\documentclass[a4paper, 12pt, fleqn]{article}
\usepackage[top=20truemm,bottom=25truemm,left=20truemm,right=20truemm]{geometry}
\usepackage[dvipdfmx]{graphicx}
\usepackage{subfigure}
\usepackage{amsmath,amssymb}
\usepackage{bm}
\usepackage{scalefnt}
\usepackage{color}
\usepackage{slashed}
\usepackage[
 colorlinks=true,
 linkcolor=magenta,
 citecolor=blue,
]{hyperref}
\usepackage{comment}
\usepackage{latexsym}
\usepackage{relsize}
\def\Babar{{\mbox{\slshape B\kern-0.1em{\smaller A}\kern-0.1em B\kern-0.1em{\smaller A\kern-0.2em R}}}}
\newcommand{\cred}[1]{{\color{red}#1}}
\newcommand{\cblu}[1]{{\color{blue}#1}}
\allowdisplaybreaks

\usepackage{authblk}

\newcommand{\Dgen}{{D^{(*)}}}
\newcommand{\Dst}{{D^*}}
\newcommand{\rDst}{r_{D^*}}
\newcommand{\rD}{r_{D}}

\newcommand{\dhh}{{\delta \hat h}}
\newcommand{\zcb}{z_{cb}}
\newcommand{\wcb}{w_{cb}}
\newcommand{\norder}[1]{{(#1)}}

\newcommand{\IWxi}[1]{\xi^{(#1)}}
\newcommand{\IWet}[1]{\eta^{(#1)}}
\newcommand{\IWci}[2]{\hat\chi^{(#1)}_{#2}}
\newcommand{\IWel}[2]{\hat\ell^{(#1)}_{#2}}

\begin{document}

\title{
\Large
Bayesian fit analysis to full distribution data of $\bar B \to \Dgen \ell\bar\nu$: 
$|V_{cb}|$~determination and New Physics constraints
}
\author[(a)]{Syuhei Iguro} 
\author[(b)]{Ryoutaro Watanabe}

\affil[(a)]{\it Department of Physics, Nagoya University, Nagoya 464-8602, Japan \vspace{0.5em}}  
\affil[(b)]{\it INFN, Sezione di Roma Tre, Via della Vasca Navale 84, 00146 Rome, Italy}

\date{August 2, 2020}

\maketitle

\begin{abstract}
We investigate the semi-leptonic decays of $\bar B \to \Dgen \ell\bar\nu$ in terms of the Heavy-Quark-Effective-Theory (HQET) parameterization for the form factors, 
which is described with the heavy quark expansion up to $\mathcal O(1/m_c^2)$ beyond the simple approximation considered in the original CLN parameterization. 
An analysis with this setup was first given in the literature, and then we extend it to the comprehensive analyses including 
(i) simultaneous fit of $|V_{cb}|$ and the HQET parameters to available experimental full distribution data and theory constraints, and 
(ii) New Physics (NP) contributions of the $V_2$ and $T$ types, such as $(\overline{c} \gamma^\mu P_Rb)(\overline{\ell} \gamma_\mu P_L \nu_{\ell})$ and $(\overline{c}  \sigma^{\mu\nu}P_Lb)(\overline{\ell} \sigma_{\mu\nu} P_L \nu_{\ell})$, to the decay distributions and rates. 
For this purpose, we perform Bayesian fit analyses by using {\tt Stan} program, a state-of-the-art public platform for statistical computation. 
Then, we show that our $|V_{cb}|$ fit results for the SM scenarios are close to the PDG combined average from the exclusive mode, and indicate significance of the angular distribution data. 
In turn, for the $\text{SM} + \text{NP}$ scenarios, our fit analyses find that non-zero NP contribution is favored at the best fit point for both $\text{SM} + V_2$ and $\text{SM} + T$ depending on the HQET parameterization model. 
A key feature is then realized in the $\bar B \to \Dgen \tau\bar\nu$ observables. 
Our fit result of the HQET parameters in the $\text{SM} (+T)$ produces a consistent value for $R_D$ while smaller for $R_\Dst$, compared with the previous SM prediction in the HFLAV report. 
On the other hand, $\text{SM}+V_2$ points to smaller and larger values for $R_D$ and $R_{D^*}$ than the SM predictions. 
In particular, the $R_\Dst$ deviation from the experimental measurement becomes smaller, which could be interesting for future improvement on measurements at the Belle~II experiment. \\

\cblu{Note added}: the present version modified typos in the formulae of Eqs.~\eqref{eq:mod1}, \eqref{eq:mod2}, \eqref{eq:mod3}, and \eqref{eq:mod4} highlighted in red, thanks to careful check by Hongkai Liu. 
We confirmed that our computation code has no such mistakes, and hence this does not affect our results. 
We also updated the attached Mathematica code for Unitarity Bound by following Ref.~\cite{Bigi:2017jbd}. 
\end{abstract}

\clearpage
\tableofcontents

\clearpage
\section{Introduction}

The semi-leptonic processes $\bar B \to \Dgen \ell\bar\nu$ for $\ell = e, \mu$ have been studied from various perspectives. 
In particular, the decay rates are of great interest as it determines the Cabibbo-Kobayashi-Maskawa~\cite{Cabibbo:1963yz,Kobayashi:1973fv} (CKM) matrix element $|V_{cb}|$ in the Standard Model (SM). 
Kinetic distributions of the processes are also important, for instance, to experimentally measure the ratios with the semi-tauonic modes, $R_{\Dgen} = \mathcal B (\bar B \to \Dgen \tau\bar\nu) / \mathcal B (\bar B \to \Dgen \ell\bar\nu)$, 
in which discrepancies between the experimental measurements and the SM predictions have been reported~\cite{Kou:2018nap}.

To investigate these issues, however, we need a sufficient knowledge on the hadron transitions $\bar B \to \Dgen$. 
In the literature, there are several theoretical descriptions on the form factors (FFs). 
The CLN parameterization~\cite{Caprini:1997mu}, applying heavy quark symmetry to FFs based on the Heavy-Quark-Effective-Theory~\cite{Isgur:1989vq,Neubert:1993mb} (HQET), has been used for this purpose. 
The BGL parameterization~\cite{Boyd:1997kz} is an alternative that relies only on QCD dispersion relations, which implies the model independent one.

An advantage of the former is that it describes the $\bar B \to D$ and $\bar B \to \Dst$ FFs with a few common parameters, and thus a combined analysis is possible, {\it e.g.}, see Ref.~\cite{Bernlochner:2017jka}.
The latter, on the other hand, includes larger number of independent parameters so that a flexible fit analysis is given, although it needs experimental data with higher statistics. 
Then, the $|V_{cb}|$ determinations from these two approaches have been in the spotlight since their results are not consistent with each other, 
see discussions in Refs.~\cite{Bigi:2016mdz,Bigi:2017njr,Grinstein:2017nlq,Jaiswal:2017rve,Bernlochner:2017xyx,Bigi:2017jbd,Gambino:2019sif}.

In the recent studies of Refs.~\cite{Jung:2018lfu,Bordone:2019vic}, the authors have revisited the HQET parameterization by adopting a setup beyond the CLN approximation and taking $1/m_c^2$ corrections into account for the heavy quark expansion. 
Then the authors have proposed two viable parameterization models introducing 13 and 23 free HQET parameters, respectively, which have to be determined from experiments and/or theoretical constraints (as also explained in this paper.)
At the expense of such a large number of parameter set, it has been found~\cite{Bordone:2019vic} that the SM fit result of $|V_{cb}|$ is in good agreement with the one obtained from the BGL parameterization. 
This conclusion, however, implies that $|V_{cb}|$ from the exclusive mode is still not fully consistent with the one from the inclusive mode, referred to as $V_{cb}$ puzzle~\cite{Tanabashi:2018oca}.

In this paper, we investigate $\bar B \to \Dgen \ell\bar\nu$ with the use of this HQET parameterization by concerning the following points: 
\begin{itemize}
 \item 
 We include all the available full distribution data of $\bar B \to \Dgen \ell\bar\nu$ from the Belle measurements~\cite{Glattauer:2015teq,Abdesselam:2017kjf,Abdesselam:2018nnh} in our fit analysis\footnote
 {The BaBar experimental analysis is given in Ref.~\cite{Dey:2019bgc}, but they do not provide detailed information on distribution data. 
 } 
 to {\it simultaneously} determine $|V_{cb}|$ and the HQET parameters. 
 Indeed, this is not the case for the reference as will be explained later. 
 \item 
 We consider New Physics (NP) effects on $\bar B \to \Dgen \ell\bar\nu$ that could affect both branching ratios and decay distributions. 
 Here, a simultaneous fit for the size of the NP contributions, $|V_{cb}|$, and the HQET parameters is performed in our analysis to see whether the $V_{cb}$ puzzle can be resolved and/or a further fit improvement is possible. 
 Then it is shown that a non-negligible NP contribution is still allowed and it satisfies the experimental data. 
 We also provide complete formulae on the decay distributions and the FFs in the presence of NP. 
 \item 
 We perform Bayesian fit analysis with the use of {\tt Stan}~\cite{Stan}, a public platform for statistical computation, which has been widely known in the statistical science community and thus could give independent check to the previous studies. 
 We also obtain quantitative evaluations on our fit results in various parameterization scenarios with/without NP by looking at {\it information criterion}~\cite{Information:Criteria}. 
\end{itemize}
In addition, we put a comparison between the CLN and HQET parameterizations to see where its difference comes out. 
Then, finally we show our predictions on the $\bar B \to \Dgen \tau\bar\nu$ observables. 
As a result, we will see that {\it NP predictions} on $R_\Dgen$ obtained from our fit results are different from the SM predictions, and that this could be a key feature for the NP search in the $\bar B \to \Dgen \tau\bar\nu$ observables.
We would like to stress that this is a comprehensive fit analysis for the HQET parameterization with/without the NP contributions.

This paper is organized as follows. 
In Sec.~\ref{Sec:theory}, we describe our theory setup for the HQET parameterization and formulae for the decay distributions in the presence of NP. 
In Sec.~\ref{Sec:fit}, we detail our fit procedure along with summary of theory constraints and experimental measurements to be taken in our analysis. 
Then we discuss our results in the various scenarios. 
Finally, a summary is put in Sec.~\ref{Sec:conclusion}. 
Details of our fit results, distribution formulae, and some theory constraints are given in Appendices.

\section{Theory setup}
\label{Sec:theory}

In this work, we start with the effective Hamiltonian that affects $\bar B \to \Dgen \ell\bar\nu$, given as 
\begin{align}
 \label{Eq:effH}
 {\mathcal H}_{\rm{eff}}=\frac{4G_F}{\sqrt2}V_{cb}
 \biggl[ (\overline{c} \gamma^\mu P_Lb)(\overline{\ell} \gamma_\mu P_L \nu_{\ell}) +C_{V_2}(\overline{c} \gamma^\mu P_Rb)(\overline{\ell} \gamma_\mu P_L \nu_{\ell}) +C_{T} (\overline{c}  \sigma^{\mu\nu}P_Lb)(\overline{\ell} \sigma_{\mu\nu} P_L \nu_{\ell}) \biggl],
\end{align} 
where $P_{L/R}=(1\mp\gamma_5)/2$ and $C_{T(V_2)} \neq 0$ indicates existence of a tensor ($V+A$ vector in $\bar c b$) type NP. 
The SM-like NP always rescales $V_{cb}$ and then we do not consider this case since its effect has to be examined by indirect or combined approaches. 
As long as the light lepton mode ($\ell = e, \mu$) is concerned, 
note that the scalar type operators, $(\overline{c}  P_Rb)(\overline{\ell} P_L \nu_{\ell})$ and $(\overline{c}  P_Lb)(\overline{\ell} P_L \nu_{\ell})$, do not affect the present processes due to the light lepton mass suppression. 
We assume that NP has $e$-$\mu$ universal ($C_X^e = C_X^\mu \equiv C_X$) and $C_X$ is real. 
This is a conservative choice since $\mathcal B (\bar B \to \Dgen \mu\bar\nu) / \mathcal B (\bar B \to \Dgen e\bar\nu) \approx 1 \pm \mathcal O(\%)$ has been reported~\cite{Glattauer:2015teq,Abdesselam:2017kjf,Abdesselam:2018nnh}. 
Also the neutrino is always taken as left-handed.

In the following part of this section, we will present theory descriptions and formulae necessary for our fit analysis.

\subsection{HQET description of Form Factors}
In the HQET basis, all possible types of the $B \to \Dgen$ current are defined as  
\begin{align}
 \langle D | \bar c \gamma^\mu b | B \rangle_\text{HQET} 
 & = \sqrt{m_Bm_D} \big[ h_+ (v+v')^\mu + h_- (v-v')^\mu \big]
 \,, \\[0.5em]
 \langle D |\bar c b| B \rangle_\text{HQET}
 & = \sqrt{m_Bm_D} (w+1) h_S
 \,, \\[0.5em]
 \langle D |\bar c \sigma^{\mu\nu} b| B \rangle_\text{HQET}
 & = -i \sqrt{m_Bm_D}\, h_T \big[ v^\mu v^{\prime\nu} - v^{\prime\mu} v^\nu  \big]
 \,, \\[0.5em]
  \langle D^* | \bar c \gamma^\mu b | B \rangle_\text{HQET} 
 & = i \sqrt{m_Bm_\Dst} h_V \varepsilon^{\mu\nu\rho\sigma} \epsilon^*_\nu v'_\rho v_\sigma 
 \,, \\[0.5em]
 \langle D^* | \bar c \gamma^\mu \gamma^5 b | B \rangle_\text{HQET} 
 & = \sqrt{m_Bm_\Dst} \big[ h_{A_1} (w+1) \epsilon^{*\mu} - (\epsilon^* \cdot v) \left( h_{A_2} v^\mu + h_{A_3} v^{\prime\mu} \right) \big] 
 \,, \\[0.5em]
  \langle D^* | \bar c \gamma^5 b | B \rangle_\text{HQET} 
 & = -\sqrt{m_Bm_\Dst} (\epsilon^* \cdot v) h_P 
 \,, \\[0.5em]
 \langle D^* | \bar c \sigma^{\mu\nu} b | B \rangle_\text{HQET} 
 & = -\sqrt{m_Bm_\Dst} \varepsilon^{\mu\nu\rho\sigma} 
 \big[ h_{T_1} \epsilon^{*}_\rho (v+v')_\sigma + h_{T_2} \epsilon^{*}_\rho (v-v')^\sigma \notag \\
 & \hspace{10em}+ h_{T_3} (\epsilon^* \cdot v) (v+v')_\rho (v-v')_\sigma \big] 
 \,, 
\end{align} 
where $v^\mu = p_B^\mu /m_B$, $v^{\prime\mu} = p_{\Dgen}^\mu /m_{\Dgen}$, $w =v \cdot v' = (m_B^2+m_{\Dgen}^2-q^2)/(2m_Bm_{\Dgen})$, and $h_X \equiv h_X(w)$ are the HQET form factors in terms of $w$. 
Then, $h_X$ can be represented by the leading Isgur-Wise~\cite{Isgur:1989vq} (IW) function $\xi$ and its correction, defined as $h_X(w) = \xi(w) \hat h_X(w)$. 
In this work, we consider 
\begin{align}
 \hat h_X = \hat h_{X,0} + {\alpha_s \over \pi} \delta \hat h_{X,\alpha_s} + {\bar \Lambda \over 2m_b} \delta \hat h_{X,m_b} + {\bar \Lambda \over 2m_c} \delta \hat h_{X,m_c} + \left({\bar \Lambda \over 2m_c}\right)^2 \delta \hat h_{X,m_c^2} \,, 
\end{align} 
where 
\begin{align}
 \hat h_{X,0} = \begin{cases} 1 & \text{for}~X = +,A_1,A_3,S,P,T,T_1 \\ 0 & \text{for}~X = -,A_2,T_2,T_3 \end{cases} \,, 
\end{align} 
and others indicate higher order corrections in $\alpha_s$ and ${1/m_{b,c}}$ expansions. 
In this work, the above HQET expansion is given at the matching scale $\mu_b = 4.2\,\text{GeV}$ with values for the expansion coefficients 
to be fixed as $\epsilon_a=\alpha_s/\pi = 0.0716$, $\epsilon_b=\bar \Lambda / (2m_b) = 0.0522$, and $\epsilon_c=\bar \Lambda / (2m_c) = 0.1807$. 
Possible uncertainties to the coefficients from quark masses are rather small, and also essentially correspond to rescaling of $\delta \hat h_{X,f}$. 
Thus we neglect those uncertainties hereafter. 
The complete expressions for $\delta \hat h_{X,f}$ are summarized in Appendix~\ref{Ap:Corrections}.

The $1/m_Q$ correction consists of three unknown sub-leading IW functions defined as $\xi_3(w)$, $\chi_2(w)$, and $\chi_3(w)$~\cite{Bernlochner:2017jka}, whereas $1/m_Q^2$ of six subsub-leading IW functions $\ell_{1\text{-}6}(w)$~\cite{Falk:1992wt}. 
Thus we have in total ten IW functions that are in principle unknown and then have to be fitted. 
We also employ the notation such as 
\begin{align}
 \label{Eq:subIW}
 \eta(w) = {\xi_3(w) \over \xi(w)} \,, 
 \quad
 \hat \chi_i(w) = {\chi_i(w) \over \xi(w)} \,, 
 \quad
 \hat \ell_i(w) = {\ell_i(w) \over \xi(w)} \,.   
\end{align}
Then, we can express any of the IW functions by means of series expansion around $w=1$. 
Namely, we take 
\begin{align}
 f(w) = \sum_{n=0} {f^{(n)} \over n!} (w-1)^n \,, 
\end{align}
for $f = \xi$, $\eta$, $\hat\chi_i$, and $\hat\ell_i$. 
Here, $f^{(n)} \equiv \left. {\partial^n f(w) \over \partial w^n} \right|_{w=1}$ are free parameters to be fitted by theoretical and/or experimental analysis. 
Analytic properties of the matrix elements indicate that the above expansion can be represented by  
\begin{align}
 w(z) = 2 \left({1+z \over 1-z}\right)^2 -1 \,, 
\end{align}
up to the order of interest. 
For instance, we have 
\begin{align}
 f(w) = f^\norder{0} + 8f^\norder{1} z + 16\left( f^\norder{1} +2f^\norder{2} \right) z^2 + {8\over3} \left(9f^\norder{1} + 48f^\norder{2} + 32f^\norder{3} \right) z^3 + \mathcal O(z^4) \,. 
\end{align}
Note that $\xi^\norder{0}=1$ and $\hat\chi_3^\norder{0}=0$ in the HQET description. 
Following Ref.~\cite{Bordone:2019vic}, the cases of 
\begin{align}
 \text{NNLO($3/2/1$)} ~:~ & \text{$\xi(w)$ up to $z^3$}\,,~~ \text{$\hat\chi_{2,3}(w)$ and $\eta(w)$ up to $z^2$} \,,~~ \text{$\hat\ell_{1\text{-}6}(w)$ up to $z^1$} \,, \\ 
 \text{NNLO($2/1/0$)} ~:~ & \text{$\xi(w)$ up to $z^2$}\,,~~ \text{$\hat\chi_{2,3}(w)$ and $\eta(w)$ up to $z^1$} \,,~~ \text{$\hat\ell_{1\text{-}6}(w)$ up to $z^0$} \,,
\end{align}
are investigated in our analysis.  
In addition, we consider  
\begin{align}
 \text{NLO($3/2/\text{-}$)} ~:~ & \text{$\xi(w)$ up to $z^3$}\,,~~ \text{$\hat\chi_{2,3}(w)$ and $\eta(w)$ up to $z^2$} \,,~~ \text{$\hat\ell_{1\text{-}6}(w) = 0$} \,, 
\end{align}
just for comparison to see how $\hat\ell_{1\text{-}6}(w)$ improves the parameter fit. 
Throughout this paper, we refer them to as the HQET parameterizations.

A final remark is that we have two kinds of expansion, namely, by $\epsilon_{a,b,c}$ and $z$ in the form factor $\hat h_X$.
A significant point is that their highest orders, as assumed above, have to be kept in observables even though it is obtained by multiplying $\hat h_X$s. 
Otherwise, higher order terms than what we take are included unfairly. 
Schematically, a proper expansion for any observable with the setup of NNLO($3/2/1$) is written as 
\begin{align}
 \label{Eq:expansion}
 \text{Obs.} = 
 &\, \mathcal O (\epsilon^0 z^0) + \mathcal O (\epsilon^0 z^1) + \mathcal O (\epsilon^0 z^2) + \mathcal O (\epsilon^0 z^3) + \mathcal O (\epsilon_a^1 z^0) + \mathcal O (\epsilon_a^1 z^1) + \mathcal O (\epsilon_a^1 z^2) + \mathcal O (\epsilon_a^1 z^3) \notag\\
 &\,+ \mathcal O (\epsilon_{b,c}^1 z^0) + \mathcal O (\epsilon_{b,c}^1 z^1) + \mathcal O (\epsilon_{b,c}^1 z^2) + \mathcal O (\epsilon_{c}^2 z^0) + \mathcal O (\epsilon_{c}^2 z^1) \,, 
\end{align}
before the $w$ integration, where $\epsilon_a = \alpha_s/\pi$ and $\epsilon_{b,c}= \bar \Lambda / (2m_{b,c})$.

\subsection{\boldmath Formula for $\bar B \to D \ell\bar\nu$: $w$ distribution}
The differential decay rate of $\bar B \to D \ell\bar\nu$ with respect to $w$ is written as 
\begin{align}
 {d\Gamma_D \over dw} = 
 G_F^2 |V_{cb}|^2 {m_Bm_D^2 \eta_\text{EW}^2 \over 48\pi^3} (1-2\rD w+\rD^2) \sqrt{w^2-1} \Big[ (1+C_{V_2})^2 H_s(w)^2 + 2|C_T|^2 H_s^T(w)^2 \Big] \,, 
\end{align}
where $\rD = m_D/m_B$ and $\eta_\text{EW} = 1.0066 \pm 0.0050$ accounts for the leading electroweak corrections~\cite{Sirlin:1981ie,Bailey:2014tva}. 
The Hadronic Amplitudes are given as~\cite{Tanaka:2012nw,Sakaki:2013bfa,Sakaki:2014sea} 
\begin{align}
 H_s(w) & = m_B \sqrt{\rD} {\sqrt{w^2-1} \over \sqrt{1-2\rD w+\rD^2}} \left[ (1+\rD) h_{+} (w) - (1-\rD) h_{-} (w) \right] \,, \notag \\[0.5em]
 H_s^T(w) & = -m_B \sqrt{\rD} \sqrt{w^2-1}\, h_{T} (w) \,. 
\end{align}
Note that the tensor NP do not interfere with the SM since the $\ell$ helicity is flipped in the massless limit of the light lepton due to spin structure of $\bar\ell \sigma^{\mu\nu} \nu$. 
One can see that  
\begin{align}
 \mathcal G(w) = h_{+} (w) - {1-\rD \over 1+\rD} h_{-} (w) \,, 
\end{align}
is the usual normalization factor for the SM. 
The FFs $h_X(w)$ are then represented with the HQET parameterizations for our analysis.

In the CLN parameterization, it is approximated with a single parameter such as $\mathcal G (w) \approx \mathcal G (1) \left[ 1- 8 \rho^2 z + (51\rho^2-10)z^2 - (252\rho^2-84)z^3 \right]$. 
Comparing it with the present forms of $h_{\pm} (w)$ for the case of $\text{NNLO} (3/2/1)$, we obtain  
\begin{align}
 \mathcal G(1)\simeq 
 &\, 1.0883 -0.1227 \IWet{0} + 0.0327 \IWel{0}{1} -0.0156 \IWel{0}{4} \,, \\[0.5em]
 -8\rho^2 \mathcal G(1) \simeq
 &\, 0.3751 +8.7061 \IWxi{1}  -7.4528 \IWci{0}{2} +22.3584 \IWci{1}{3} -0.9816 \big( \IWet{1} + \IWet{0} \IWxi{1} \big) \notag \\
 &\,  +0.2612 \big( \IWel{1}{1} + \IWel{0}{1} \IWxi{1} \big) -0.1247 \big( \IWel{1}{4} + \IWel{0}{4} \IWxi{1} \big) \,, 
\end{align}
in our setup. 
We can see that the NNLO parameters $\IWel{n}{1,4}$ affect these quantities. 
As for the $z^2$ and $z^3$ terms, the HQET parameterizations take lengthy forms while the CLN gives the above approximate expressions by the single parameter $\rho^2$. 
We will show the latter approximation underestimates uncertainties in these terms.

\subsection{\boldmath Formula for $\bar B \to D^* \ell\bar\nu$: full angular distribution}
Concerning the available experimental data, we show the full differential decay rate for $B^0 \to \Dst^- (\to \bar D^0 \pi^-) \ell\bar\nu$ in the presence of the NP contributions: 
\begin{align}
 \label{eq:fulldistribution}
 {d\Gamma^\text{full}_\Dst \over dw\,d\cos\theta_\ell\,d\cos\theta_V\,d\chi} = 
 &\, \mathcal B(\Dst^- \to \bar D^0 \pi^-) G_F^2 |V_{cb}|^2 {3m_Bm_\Dst^2\eta_\text{EW}^2 \over 4 (4\pi)^4} \\
 &\, \times (1-2\rDst w+\rDst^2) \sqrt{w^2-1} \sum_{i=1}^6 \mathcal J_i(\theta_\ell,\theta_V,\chi) \mathcal H_i (w) \,, \notag
\end{align}
where $\mathcal J_i$ include angular dependences\footnote{
Note that the definition of $\theta_\ell$ here is not the same as $\theta_\tau$ in Ref.~\cite{Tanaka:2012nw}, but related as $\theta_\ell = \pi -\theta_\tau$. 
} 
obtained as 
\begin{align}
 \mathcal J_1 &= (1-\cos\theta_\ell)^2\sin^2\theta_V \,, & \mathcal J_2 &= (1+\cos\theta_\ell)^2\sin^2\theta_V \,, \notag \\ 
 \mathcal J_3 &= 4\sin^2\theta_\ell \cos^2\theta_V \,  & \mathcal J_4 &= -2\sin^2\theta_\ell\sin^2\theta_V\cos2\chi \,, \notag \\ 
 \mathcal J_5 &= -4\sin\theta_\ell(1-\cos\theta_\ell)\sin\theta_V\cos\theta_V\cos\chi \,, & & \notag \\
 \mathcal J_6 &= +4\sin\theta_\ell(1+\cos\theta_\ell)\sin\theta_V\cos\theta_V\cos\chi\,, & & \notag \\
 \mathcal J_7 &= 2 \sin^2\theta_\ell \sin^2\theta_V \,, & \mathcal J_8 &= 8 \cos^2\theta_\ell \cos^2\theta_V \,, \label{eq:angular}
\end{align}
and $\mathcal H_i$ indicate hadronic parts described as 
\begin{align}
 \mathcal H_1(w) & =  \Big( H_+(w) -C_{V_2} H_-(w) \Big)^2 \,, \notag \\
 \mathcal H_2(w) & = \Big( H_-(w) -C_{V_2} H_+(w) \Big)^2 \,, \notag \\[0.5em]
 \mathcal H_3(w) & = (1-C_{V_2})^2 H_0(w)^2 \,, \notag \\[0.5em]
 \mathcal H_4(w) & = \Big( H_+(w) -C_{V_2} H_-(w) \Big) \Big( H_-(w) -C_{V_2} H_+(w) \Big) + 16|C_T|^2 H^T_-(w) H^T_+(w) \,, \notag \\
 \mathcal H_5(w) & = (1-C_{V_2}) H_0(w) \Big( H_+(w) -C_{V_2} H_-(w) \Big) +8|C_T|^2 \Big( H^T_0(w) H^T_-(w) -H^T_0(w) H^T_+(w) \Big) \,, \notag \\
 \mathcal H_6(w) & = (1-C_{V_2}) H_0(w) \Big( H_-(w) -C_{V_2} H_+(w) \Big) +8|C_T|^2 \Big( H^T_0(w) H^T_-(w) -H^T_0(w) H^T_+(w) \Big) \,, \notag \\
 \mathcal H_7(w) & = 8|C_T|^2 \Big( H^T_+(w)^2 + H^T_-(w)^2 \Big) \,, \notag \\[0.5em]
 \mathcal H_8(w) & = 8|C_T|^2 H^T_0(w)^2\,. 
\end{align}
Note again that there is no interference term between the vector and tensor currents. 
Then, we can write the Hadronic Amplitudes $H_n^{(T)}$ from Refs.~\cite{Tanaka:2012nw,Sakaki:2013bfa,Sakaki:2014sea} as 
\begin{align}
 H_\pm(w) & = m_B \sqrt{\rDst} \left[ (w+1) h_{A_1} (w) \mp \sqrt{w^2-1} h_{V} (w) \right] \,, \notag \\[0.5em] 
 H_0(w) & = m_B \sqrt{\rDst} {\cred{w+1} \over \sqrt{1-2\rDst w+\rDst^2 }} \Big[ (\rDst - w) h_{A_1} (w) +(w-1) (\rDst h_{A_2} (w) + h_{A_3} (w)) \Big] \,, \notag \\[0.5em]
 H^T_\pm(w) & = \cred{\pm} m_B \sqrt{\rDst} {1-\rDst (w \mp \sqrt{w^2-1}) \over \sqrt{1-2\rDst w+ \rDst^2 }} \notag \\
 &\hspace{6em}\times \left[ h_{T_1} (w) + h_{T_2} (w) + (w \pm \sqrt{w^2-1}) (h_{T_1} (w) - h_{T_2} (w)) \right] \,, \notag \\[0.5em]
 H^T_0(w) & = -m_B \sqrt{\rDst} \Big[ (w+1) h_{T_1} (w) +(w-1) h_{T_2} (w) +2(w^2-1) h_{T_3} (w) \Big] \,. \label{eq:mod1}
\end{align}
The angular dependence of Eq.~\eqref{eq:angular} can be derived as explained in Appendix~\ref{Ap:AngularFormula}. 
The normalization factor is given as $\mathcal F(1) = h_{A_1}(1)$ and then the HQET parameterization leads to 
\begin{align}
 \mathcal F(1)\simeq 0.9702 + 0.0327 \IWel{0}{2} \,, 
\end{align}
and its $w$ dependence takes a lengthy form, (and hence omitted.)

In the CLN parameterization, on the other hand, the $w$ dependence on $\mathcal F(w)$ is approximated by using the following functions: 
$h_{A_1}(w)$ with the slope $\rho^2_{D^*}$ similar to $\mathcal G(w)$, $R_1(w) \equiv h_{V}(w)/h_{A_1}(w)$, and $R_2(w) \equiv \big( h_{A_3}(w) + r_\Dst h_{A_2}(w) \big) /h_{A_1}(w)$, where the latter two are expanded in $(w-1)$. 
But, this simplification is not proper for the present parameterization since it conflicts with the $\epsilon_{a,b,c}$ and $z$ expansions which we explained as in Eq.~\eqref{Eq:expansion}. 
Instead, we will provide a more straightforward fit result by means of the $z$ expansion for $\mathcal F(w)$.

\section{Fit analysis}
\label{Sec:fit}

\subsection{Theory constraints on form factors}
There are theoretical studies to evaluate the form factors at specific points of $w$ with respect to the following quantities: 
\begin{align}
  f_+^{B \to D} (w) & = {1 \over 2 \sqrt{\rD}} \Big[ \left(1+r_D\right) h_+ (w) - \left(1-r_D\right) h_-(w) \Big] \,, \\
  f_0^{B \to D} (w) & = \sqrt{\rD} \left[ {w + 1 \over 1+\rD} h_+ (w) + {w - 1 \over 1-\rD} h_- (w) \right] \,, \\ 
  f_T^{B \to D} (w) & = {1+\rD \over 2 \sqrt{\rD}} h_T (w) \,, \\
  A_1^{B \to D^*} (w) & = {\sqrt{\rDst}(1+w) \over 1+\rDst} h_{A_1} (w) \,, \\
   A_0^{B \to D^*} (w) & = {1 \over 2 \sqrt{\rDst}} \Big[ (w+1) h_{A_1} (w) + (w\, \rDst - 1) h_{A_2} (w) +(\rDst -w)h_{A_3} (w) \Big] \,, \\
  V^{B \to D^*} (w) & = {1+r_\Dst \over  2\sqrt{r_\Dst}} h_V (w) \,, \\
  T_1^{B \to D^*} (w) & = {1 \over 2\sqrt{r_\Dst}} \Big[ (1+r_\Dst)h_{T_1} (w) - (1-r_\Dst)h_{T_2} (w) \Big] \,, \\
  T_2^{B \to D^*} (w) & = \sqrt{\rDst} \left[ {w+1 \over 1+\rDst} h_{T_1} (w) - {w-1 \over 1-\rDst} h_{T_2} (w) \right] \,, \\
  T_{23}^{B \to D^*}(w) & = {1+r_\Dst \over 4 \sqrt{r_\Dst}} \Big[ (w+1) h_{T_1} (w) + (w-1) h_{T_2} (w) - (w^2-1) h_{T_3} (w) \Big] \,. 
\end{align} 
Then, the lattice studies~\cite{Lattice:2015rga,Na:2015kha,Aoki:2019cca} provide the following evaluations  
\begin{align}
  \label{Eq:lattice1}
  f_+^{B \to D} (\{1,~ 1.08,~ 1.16\}) & = \{1.1994(95),~ 1.0941(104),~ 1.0047(123) \} \,, \\
  f_0^{B \to D} (\{1,~ 1.08,~ 1.16\}) & = \{ 0.9026(72),~ 0.8609(77),~ 0.8254(94) \} \,, \\
  \label{Eq:lattice3}
  h_{A_1} (1) &= 0.904 \pm 0.012 \,. 
\end{align} 
In Ref.~\cite{Gubernari:2018wyi}, the form factors at $q^2=\{0, -5, -10, -15 \} \text{GeV}^2$ have been evaluated by a light-cone sum rule (LCSR) approach. 
The result can be summarized as 
\begin{align}
 \label{Eq:LCSR1}
 f_+^{B \to D} (\{1.59,~ 1.84,~ 2.10,~ 2.35 \}) & = \{ 0.65(8),~ 0.55(7),~ 0.48(6),~ 0.42(5) \} \,, \\
 f_0^{B \to D} (\{1.84,~ 2.10,~ 2.35 \}) & = \{ 0.62(8),~ 0.59(7),~ 0.56(7) \} \,, \\
 f_T^{B \to D} (\{1.59,~ 1.84,~ 2.10,~ 2.35 \}) & = \{ 0.57(5),~ 0.46(3),~ 0.38(3),~ 0.33(2) \} \,, \\
 A_1^{B \to D^*} (\{1.50,~ 1.74,~ 1.98,~ 2.21 \}) & = \{ 0.60(9),~ 0.56(9),~ 0.53(9),~ 0.50(8) \} \,, \\
 A_0^{B \to D^*} (\{1.74,~ 1.98,~ 2.21 \}) & = \{ 0.55(8),~ 0.47(7),~ 0.40(7) \} \,, \\
 V^{B \to D^*} (\{1.50,~ 1.74,~ 1.98,~ 2.21 \}) & = \{ 0.69(13),~ 0.59(11),~ 0.50(9),~ 0.44(8) \} \,, \\
 T_1^{B \to D^*} (\{1.50,~ 1.74,~ 1.98,~ 2.21 \}) & = \{ 0.63(10),~ 0.54(9),~ 0.45(8),~ 0.39(7) \} \,, \\
 T_2^{B \to D^*} (\{1.74,~ 1.98,~ 2.21 \}) & = \{ 0.60(10),~ 0.58(10),~ 0.55(10) \} \,, \\
 T_{23}^{B \to D^*} (\{1.50,~ 1.74,~ 1.98,~ 2.21 \}) & = \{ 0.81(11),~ 0.74(11),~ 0.69(10),~ 0.65(10) \} \,, 
 \label{Eq:LCSR9}
\end{align}
where $w = \{1.59, \cdots \}$ and $w = \{1.50, \cdots \}$ correspond to $q^2=\{0, \cdots\}$ for $B \to D$ and $B \to D^*$, respectively. 
Thanks to this comprehensive work, for instance, a fit analysis to ``theory constraints only'' is even possible.

In addition, QCD sum rule (QCDSR) can evaluate the sub-leading IW functions as in Refs.~\cite{Neubert:1992wq,Neubert:1992pn,Ligeti:1993hw}. 
By using formulae in the literature with updated QCD input data, we derive the following constraints 
\begin{align}
 \label{Eq:QCDSR1}
 -0.08 & < \IWci{0}{2} <-0.04 \,, & -0.02 & < \IWci{1}{2} <+0.02 \,, & -0.03 & < \IWci{2}{2}  <+0.01 \,, \\
 & & +0.01 & < \IWci{1}{3} <+0.06 \,, & -0.07 & < \IWci{2}{3} <+0.02 \,, \\
 +0.50 & < \IWet{0} <+0.73 \,, & +0.01 & < \IWet{1} <+0.07 \,, & -0.12 & < \IWet{2} <-0.02 \,. 
 \label{Eq:QCDSR3}
\end{align}
We show a detail of these constraints in Appendix \ref{Ap:QCDSR}.

In addition, we need to take care of Unitarity Bound (UB) for the case of the HQET parameterization. 
Following Eqs.(5) -- (20) of Ref.~\cite{Caprini:1997mu}, we obtain the functions $U_{J^P}$ in terms of the present HQET parameters to be constrained by  
\begin{align}
 \label{Eq:UB1}
 U_{0^+} & < \chi_{0^+} (0) \approx (5.96 \pm 0.44) \times10^{-3} \,, \\
 U_{0^-} & < \tilde\chi_{0^-} (0) \approx (1.33 \pm 0.04) \times10^{-2} \,, \\
 U_{1^+} & < m_{B^*}m_{D^*} \chi_{1^+} (0) \approx (3.90 \pm 0.16) \times10^{-3} \,, \\
 U_{1^-} & < m_{B^*}m_{D^*} \tilde\chi_{1^-} (0) \approx (3.58 \pm 0.18) \times10^{-3} \,, 
 \label{Eq:UB4}
\end{align}
where the explicit forms of $U_{J^P}$ are a bit lengthy and thus we put a {\tt Mathematica} file in the source of \href{https://arxiv.org/abs/2004.10208}{the ArXiv version}. 
The above numerical bounds are obtained by using recent data of (excited) $B_c$ states~\cite{Eichten:2019gig,Li:2019tbn} and quark masses~\cite{Tanabashi:2018oca}, instead of the original one~\cite{Caprini:1997mu}.

For now, we leave discussion on how we take the uncertainties of these theoretical constraints in our fit analysis. 
It will be explained in Sec.~\ref{Sec:procedure}, and discussed in Sec.~\ref{Sec:thuncertainty}.

\subsection{Experimental data}
The kinetic distributions of $\bar B \to \Dgen \ell\bar\nu$ have been measured by the Belle collaboration in Refs.~\cite{Glattauer:2015teq,Abdesselam:2017kjf,Abdesselam:2018nnh}. 
Available experimental data are then $w$ distributions of $\bar B \to D \ell\bar\nu$~\cite{Glattauer:2015teq} (denoted as Belle15), 
and full kinetic $(w, \theta_\ell, \theta_V, \chi)$ distributions of $\bar B \to D^* \ell\bar\nu$ with the successive decay $D^* \to D\pi$~\cite{Abdesselam:2017kjf,Abdesselam:2018nnh}. 
The latter includes two independent measurements; one with hadronic tagging~\cite{Abdesselam:2017kjf} (Belle17) and with untagged approach~\cite{Abdesselam:2018nnh} for each $e$ and $\mu$ mode (Belle18-$e$ and Belle18-$\mu$).

The Belle15 data correspond to the binned decay rate with respect to $w$, where the four processes, $\bar B^0 \to D^+e^-\bar\nu$, $\bar B^0 \to D^+\mu^-\bar\nu$, $\bar B^- \to D^0e^-\bar\nu$, and $\bar B^- \to D^0\mu^-\bar\nu$, are combined. 
The Belle17 data are given in terms of the unfolded decay rate of $\bar B^0 \to D^{*+} \ell^-\bar\nu$ for a corresponding bin ${\Delta \Gamma \over\Delta x}$. 
This is derived from Eq.~\eqref{eq:fulldistribution} as 
\begin{align}
 {\Delta \Gamma \over \Delta x} = {1 \over \mathcal B(\Dst^+ \to D^0 \pi^+)}  \int_{\Delta x} {d\Gamma^\text{full}_\Dst \over dx} \,, 
\end{align}
for $x = (w, \cos\theta_\ell, \cos\theta_V, \chi)$. 
On the other hand, the Belle18 data are shown in terms of binned signal event $\left. {\Delta N \over \Delta x} \right|_i$ (for $i$-th bin) in which the folded effect is presented by Response Matrix $\mathcal R$ together with efficiency $\varepsilon$ among the bins. 
This is obtained as 
\begin{align}
 \left. {\Delta N \over \Delta x} \right|_i = N_{B^0}\, \tau_{B^0}\, \mathcal B(\bar D^0 \to K^- \pi^+)\, \mathcal R_{ij}\, \varepsilon_j \int_{\Delta x_j} {d\Gamma^\text{full}_\Dst \over dx} \,, 
\end{align}
where $N_{B^0}$, $\mathcal R_{ij}$, and $\varepsilon_j$ are provided in Ref.~\cite{Abdesselam:2018nnh} for each $e$ and $\mu$ modes.

Furthermore, we also take the world averages of the branching ratios (BR) of $\bar B \to \Dgen \ell\bar\nu$~\cite{Tanabashi:2018oca} in our fit.
A short summary for the experimental data and the theory constraints is shown in Table~\ref{Tab:datasummary}. 
Correlations among the bins for each measurement are also taken into account in our fit analysis, (see corresponding references.) 
%
\begin{table}[t]
\renewcommand{\arraystretch}{1.3}
\begin{center}
\scalebox{0.95}{
\begin{tabular}{lll}
 \hline\hline
 \multicolumn{1}{l}{Name} & Object & Description \\
 \hline
 ~Belle15~\cite{Glattauer:2015teq} & ~$\bar B \to D \ell\bar\nu$ & ~$w$ distribution (10) \\
 \hline
 ~Belle17~\cite{Abdesselam:2017kjf} & ~$\bar B^0 \to D^{*+} \ell^-\bar\nu$ & ~$(w, \theta_\ell, \theta_V, \chi)$ distributions $(40)$ \\
 \hline
 ~Belle18-$e$~\cite{Abdesselam:2018nnh} & ~$\bar B^0 \to D^{*+} e^-\bar\nu$ & ~$(w, \theta_\ell, \theta_V, \chi)$ distributions $(40)$  \\
 \hline
 ~Belle18-$\mu$~\cite{Abdesselam:2018nnh} & ~$\bar B^0 \to D^{*+} \mu^-\bar\nu$ & ~$(w, \theta_\ell, \theta_V, \chi)$ distributions $(40)$  \\
 \hline
 ~BR~\cite{Tanabashi:2018oca} & ~$\bar B \to \Dgen \ell\bar\nu$ & ~branching ratios $(2)$ \\
 \hline
 ~Lattice~\cite{Lattice:2015rga,Na:2015kha,Aoki:2019cca} & ~FFs $(f_{+,0}^{B \to D}, h_{A_1})$ & ~Eqs.~\eqref{Eq:lattice1}--\eqref{Eq:lattice3} constraints $(7^*)$ \\
 \hline
 ~LCSR~\cite{Gubernari:2018wyi} & ~FFs $(f_{+,0,T}^{B \to D}, A_{1,0}^{B \to D^*}, V^{B \to D^*}, T_{1,2,23}^{B \to D^*} )$ & ~Eqs.~\eqref{Eq:LCSR1}--\eqref{Eq:LCSR9} constraints $(33)$ \\ 
 \hline
 ~QCDSR~\cite{Neubert:1992wq,Neubert:1992pn,Ligeti:1993hw} & ~FFs $(\IWci{n}{2,3}, \IWet{n})$ & ~Eqs.~\eqref{Eq:QCDSR1}--\eqref{Eq:QCDSR3} constraints $(8)$ \\ 
 \hline
 ~UB~\cite{Caprini:1997mu} & ~$U_{J^P}$ (see the attached file) & ~Eqs.~\eqref{Eq:UB1}--\eqref{Eq:UB4} constraints $(4)$ \\
 \hline\hline
\end{tabular}
} 
\caption{   
Summary of the experimental data and the theory constraints used in our fit analysis. 
Numbers of independent data points are also exhibited in brackets. 
$(^*)$ The relation $f_+(q^2\!=\!0) = f_0(q^2\!=\!0)$ implies that the lattice result has only 6 independent observables. 
}
\label{Tab:datasummary}
\end{center}
\end{table}

\subsection{Fit procedure}
\label{Sec:procedure}
In this work, a Bayesian fit analysis is applied to obtain allowed ranges of the HQET parameters and $|V_{cb}|$ with the use of Markov-Chain-Monte-Carlo (MCMC) method by {\tt Stan}~\cite{Stan}, 
{\it a state-of-the-art platform for statistical modeling and high-performance statistical computation}, implemented in {\tt MathematicaStan}~\cite{MathStan}. 
The analysis is performed by MCMC runs involving 10 chains with Hamiltonian Monte Carlo algorithm giving $10^4$ sampling points for every fit.

Although {\tt Stan} is widely known in the statistical science community, it has not often been used in particle physics analysis. 
This enables us to give independent check of fit results obtained from public/private codes developed by particle physicists.

Our fit procedure is briefly exhibited as follows. 
We basically take into account the full experimental data points of $\bar B \to \Dgen \ell\bar\nu$ and the applicable theoretical constraints on the specific FFs, as summarized in Table~\ref{Tab:datasummary}. 
Namely, 184 data points are used to fit the free parameters. 
Regarding the theory constraints, we need to declare ways of treating uncertainties. 
First, we simply take them as normal distributions in order to obtain mean values and variances from the sampling points of the fitted parameters. 
As for the UBs, it is assumed such as, {\it e.g.}, $(U_{0^+} - 0)^2/\chi_{0^+}(0)^2$. 
This means that $1\sigma$ deviation is the threshold for UB which should have to be satisfied in a final result. 
We will check this point later in Sec.~\ref{Sec:thuncertainty}. 
After then, we will also discuss the QCDSR bounds on $\IWci{n}{2,3}$ and $\IWet{n}$ since they include special input of $T$ and $\omega_0$ (see Appendix~\ref{Ap:QCDSR} for more detail) that have no fair description of ``central value''.\footnote{
This issue might be similar for the LCSR bounds, but it is beyond the scope of the present work.  
}

For comparison and later discussion, we also consider the following case where limited data points are taken into account for a fit analysis: 
{\bf\boldmath $w+$theory} -- only the $w$ distributions along with the theory constraints and the branching ratios.

As for the phenomenological mode, we investigate SM, $\text{SM}+V_2$, $\text{SM}+T$, and $\text{SM}+V_2+T$ as described in Eq.~\eqref{Eq:effH} with the HQET parameterization for the FFs. 
Then we evaluate {\it Information Criterion} that offers the predictive accuracy of the model. 
To be precise, we employ cAIC defined as~\cite{Information:Criteria} 
\begin{align}
 \text{cAIC}  = -2 \ln \mathcal L + {2k (k+1) \over n-k-1} \,, 
\end{align}
where $\mathcal L$ is the maximum likelihood and $n$ ($k$) denotes the number of data points (the model parameters to be fitted). 
The second term gives a penalty for overestimate of increasing number of model parameters.  
In our case, $k = 23+1 (+1 \,\,\text{or}\, +2)$ in the $\text{SM} (+\text{NP})$ for NNLO ($3/2/1$), and similarly $k=13+1(+1 \,\,\text{or}\, +2)$ for ($2/1/0$), with $n = 184$.
A preferred model has a smaller cAIC.

\subsection{Result}
First, we show our fit results of the HQET parameters and $|V_{cb}|$ for the $\text{SM} (+ \text{NP})$ scenarios at the NNLO heavy quark expansion in Table~\ref{Tab:fitresult}. 
We also evaluate how the present phenomenological models improve fit to data points by looking at difference in {\it Information Criterion} from a reference model. 
We define $\Delta \text{IC}_\text{model} = \text{cAIC}_0 - \text{cAIC}_\text{model}$, where $\text{cAIC}_0 = 987.4$ is the fit result of our reference model, SM NLO($3/2/\text{-}$). 
We remark that a larger value of $\Delta \text{IC}$ implies a better improvement from the reference model. 
As seen from the result, all the present models improve the fit compared with the reference model in which $\IWel{n}{i} = 0$ is taken. 
This illustrates significance of non-zero values (beyond variances) for the NNLO parameters.

On the other hand, one finds that SM ($2/1/0$) is more preferred than SM ($3/2/1$). 
This is also similar to the cases of the $\text{SM} + \text{NP}$ scenarios, whose details will be shown in the following subsection.
This means that 23 HQET parameters in ($3/2/1$) are surplus to the present available experimental/theory 184 data points, and then 13 in ($2/1/0$) are sufficient to explain the available data points at present. 
However, we believe that this is not conclusive since it could vary as additional measurements become available in the future, {\it e.g.}, by the Belle~II experiment. 
Therefore, we still continue to examine the both cases of ($2/1/0$)  and ($3/2/1$) in the following part of this work. 
For more details of our fit results, such as correlation matrix, see Appendix~\ref{Ap:FitResult}.

\begin{table}[t!]
\renewcommand{\arraystretch}{1.3}
  \begin{center}
  \scalebox{0.8}{
  \begin{tabular}{|c||c|c||c|c|c|c|}	
  \hline\hline
  FF scenario & \multicolumn{2}{|c||}{$(3/2/1)$} & \multicolumn{4}{|c|}{$(2/1/0)$} \\ 
  \hline
 Model   & SM & $\text{SM}\!+\!V_2$ & SM & $\text{SM}\!+\!V_2$ & $\text{SM}\!+\!T$ & $\text{SM}\!+\!V_2+\!T$ \\ 
  \hline
  $|V_{cb}| \times 10^3$ & $39.3 \pm 0.6$  & $39.9 \pm 0.5$	& $39.7 \pm 0.6$ 	& $39.9 \pm 0.6$		& $39.7 \pm 0.6$		& $39.9 \pm 0.6$ \\
  \hline
  $C_\text{NP}$ 	     & - 			& $0.05 \pm 0.01$	& -			  	& $0.02 \pm 0.01$ 		& $|0.02 \pm 0.01|$		& \begin{tabular}{c} $V_2: 0.02 \pm 0.01$ \\ $T: |0.02 \pm 0.01|$ \end{tabular}  \\  
  \hline\hline
  $\IWxi{1}$  		& $-0.93 \pm 0.10$   & $-0.94 \pm 0.09$	& $-1.10 \pm 0.04$	& $-1.09 \pm 0.04$	 	&  $-1.09 \pm 0.04$		& $-1.09 \pm 0.04$    \\
  \hline
  $\IWxi{2}$  		& $+1.35 \pm 0.26$  &  $+1.37 \pm 0.25$	& $+1.57 \pm 0.10$	& $+1.55 \pm 0.10$ 		& $+1.56 \pm 0.10$		& $+1.54 \pm 0.10$  \\
  \hline
  $\IWxi{3}$  		& $-2.67 \pm 0.75$   & $-2.71 \pm 0.73$	& -	   			& -					& -  					& - \\
  \hline\hline
  $\IWci{0}{2}$  		& $-0.05 \pm 0.02$   & $-0.05 \pm 0.02$	& $-0.06 \pm 0.02$	& $-0.06 \pm 0.02$		& $-0.06 \pm 0.02$		& $-0.06 \pm 0.02$ \\
  \hline
  $\IWci{1}{2}$  		& $+0.01 \pm 0.02$  & $+0.01 \pm 0.02$	& $+0.01 \pm 0.02$	& $+0.01 \pm 0.02$		& $+0.01 \pm 0.02$		& $+0.00 \pm 0.02$   \\
  \hline
  $\IWci{2}{2}$  		&  $-0.01 \pm 0.02$	& $-0.02 \pm 0.02$	& -  				& -					& -					& -  \\
  \hline
  $\IWci{1}{3}$  		& $-0.05 \pm 0.02$   & $-0.05 \pm 0.02$	& $-0.03 \pm 0.01$   & $-0.04 \pm 0.01$		& $-0.03 \pm 0.01$		& $-0.04 \pm 0.01$ \\
  \hline
  $\IWci{2}{3}$  		& $-0.03 \pm 0.03$	& $+0.01 \pm 0.03$	& -  				& -					&- 					& -  \\
  \hline
  $\IWet{0}$  		& $+0.74 \pm 0.11$  	& $+0.71 \pm 0.11$	& $+0.38\pm0.06$	& $+0.37\pm0.06$		& $+0.40\pm0.06$		& $+0.38\pm0.06$  \\
  \hline
  $\IWet{1}$  		& $+0.05\pm 0.03$ 	& $+0.05\pm 0.03$	& $+0.08\pm0.03$    & $+0.08\pm0.03$		& $+0.08\pm0.03$ 		& $+0.07\pm0.03$\\
  \hline
  $\IWet{2}$  		& $-0.05 \pm 0.05$	& $-0.06 \pm 0.05$	& -    			& -					& - 					& -   \\
  \hline\hline
  $\IWel{0}{1}$  		& $+0.09 \pm 0.18$ 	& $+0.19 \pm 0.18$	& $+0.50 \pm 0.16$ 	& $+0.48 \pm 0.16$		& $+0.50 \pm 0.16$		& $+0.49 \pm 0.16$  \\ 
  \hline
  $\IWel{1}{1}$  		&  $+1.20 \pm 2.09$  & $-0.70 \pm 1.92$	& -  				& -					& -					& -  \\ 
  \hline
  $\IWel{0}{2}$  		&  $-2.29 \pm 0.33$ 	& $-1.64 \pm 0.36$	& $-2.16 \pm 0.29$	& $-1.93 \pm 0.32$		& $-2.24 \pm 0.29$		&  $-2.00 \pm 0.33$  \\ 
  \hline
  $\IWel{1}{2}$  		& $-3.66 \pm 1.56$	& $-2.92 \pm 1.55$	& -				& -					& - 					& -   \\ 
  \hline
  $\IWel{0}{3}$  		& $-1.90 \pm 12.4$  	& $-1.50 \pm 12.6$	& $-1.14 \pm 2.34$	& $-0.23 \pm 2.39$		& $-1.21 \pm 2.29$		&  $-0.32 \pm 2.41$  \\ 
  \hline
  $\IWel{1}{3}$  		& $+3.91 \pm 4.35$  	& $+4.29 \pm 4.31$	& -				& -					& -					& -  \\ 
  \hline
  $\IWel{0}{4}$  		& $-2.56 \pm 0.94$ 	& $-2.22 \pm 0.94$	& $+0.82 \pm 0.47$ 	& $+0.97 \pm 0.48$		& $+0.76 \pm 0.47$		& $+0.94 \pm 0.49$  \\ 
  \hline
  $\IWel{1}{4}$  		& $+1.78 \pm 0.93$  	& $+1.82 \pm 0.91$	& -				& -					& - 					& - \\ 
  \hline
  $\IWel{0}{5}$  		& $+3.96 \pm 1.17$  	& $+6.31 \pm 1.32$	& $+1.39 \pm 0.43$ 	&  $+2.03 \pm 0.59$ 		& $+1.32 \pm 0.43$		& $+1.99 \pm 0.59$  \\ 
  \hline
  $\IWel{1}{5}$  		& $+2.10 \pm 1.47$ 	& $+2.29 \pm 1.51$	& -				& -					& -					& -  \\ 
  \hline
  $\IWel{0}{6}$  		& $+4.96 \pm 5.76$ 	& $+7.15 \pm 5.87$	& $+0.17 \pm 1.15$	& $+0.90 \pm 1.23$		& $+0.06 \pm 1.15$		&  $+0.81 \pm 1.24$	   \\ 
  \hline
  $\IWel{1}{6}$  & 	$+5.08 \pm 2.97$ 	& $+5.52 \pm 3.04$	& -				& -					& - 					& -   \\ 
  \hline\hline
  $\Delta \text{IC}$ & $118.1$ 			& $128.4$			& $162.4$			& $161.5$				& $161.3$				& $160.4$  \\
  \hline\hline
  \end{tabular} 
  }
  \caption{   
  Fit results of the simultaneous determinations for the HQET parameters and $|V_{cb}|$ in several phenomenological models at NNLO with/without NP. 
  Larger value of $\Delta \text{IC}$ indicates better improvement of the fit from the reference model of NLO($3/2/\text{-}$). 
  }
  \label{Tab:fitresult}
  \end{center}
\end{table}

As a consistency check with integrated observables, we generate the branching ratios and the $D^*$ polarization ($e$ mode) that result in 
\begin{align}
 \mathcal B (\bar B^0 \to D^+ \ell^-\bar\nu)_\text{SM} & = \Big[ (2.23 \pm 0.05) \% \,; ~ (2.20 \pm 0.05) \%  \Big] \,, \\ 
 \mathcal B (\bar B^0 \to D^{*+} \ell^- \bar\nu)_\text{SM} & = \Big[ (5.06 \pm 0.05) \% \,; ~ (5.07 \pm 0.05) \%  \Big] \,, \\ 
 F_L^{D^*} (\bar B^0 \to D^{*+} e^- \bar\nu)_\text{SM} & = \Big[ 0.534 \pm 0.002 \,; ~ 0.531 \pm 0.002 \Big] \,, 
\end{align}
in the SM for the cases of $\big[(2/1/0);~ (3/2/1) \big]$, respectively.
This is compared with the experimental measurements of $\mathcal B (\bar B^0 \to D^+ \ell^-\bar\nu)_\text{exp} = (2.31 \pm 0.03 \pm 0.09) \%$ and $ \mathcal B (\bar B^0 \to D^{*+} \ell^- \bar\nu)_\text{exp} = (5.06 \pm 0.02 \pm 0.12) \%$ from the world average~\cite{Amhis:2019ckw}, 
and $F_L^{D^*} (\bar B^0 \to D^{*+} e^- \bar\nu)_\text{exp} = 0.56 \pm 0.02$ from Ref.~\cite{Abdesselam:2019wbt}. 
Note that $F_L^{D^*} (\bar B^0 \to D^{*+} e^- \bar\nu)_\text{exp}$ is still preliminary (and thus we did not take it in our fit.) 
We can see that they are in good agreements within uncertainties, but the best fit point for the $D$ mode is a bit smaller than data, even though it was included in the fit.

\subsubsection{$|V_{cb}|$ determination}
Our fit results for $|V_{cb}|$ in the SM $(2/1/0)$ and $(3/2/1)$ scenarios are both close to the PDG combined average, $(39.5 \pm 0.9) \times 10^{-3}$, from the exclusive mode~\cite{Tanabashi:2018oca}. 
In Table~\ref{Tab:comparisonVcb}, we put summary for the recent $|V_{cb}|$ determinations along with the normalization factors $\mathcal G(1)$ and $\mathcal F(1)$.

Here we would like to discuss difference in the $|V_{cb}|$ determination between our results and one from Ref.~\cite{Bordone:2019vic}. 
In their work, $|V_{cb}|$ has been extracted by using the fit result of the HQET parameters, and after then, by taking the integrated branching ratios of $\bar B \to \Dgen \ell\bar\nu$. 
Although the former fit analysis includes the experimental $w$ distributions, it is utilized only to fit the HQET parameters. 
Indeed, we find that their result can be reproduced when we perform the fit analysis with the data set of {\bf\boldmath $w+$theory} as shown in Table~\ref{Tab:comparisonVcb}. 
Therefore, we emphasize that the angular distributions are also significant for the $|V_{cb}|$ determination.

\begin{table}[t]
  \renewcommand{\arraystretch}{1.3}
  \begin{center}
  \scalebox{0.85}{
  \begin{tabular}{c|ccc|cc}
  \hline\hline
   & {\bf all} $(2/1/0)$ & {\bf all} $(3/2/1)$ & PDG/HFLAV~\cite{Tanabashi:2018oca,Amhis:2019ckw} & {\bf\boldmath $w+$theory} $(3/2/1)$ & Ref.~\cite{Bordone:2019vic} $(3/2/1)$ \\ 
  \hline
  $|V_{cb}| \times 10^3$ 
  & $39.7 \pm 0.6$ 
  & $39.3 \pm 0.6$ 
  & $39.5 \pm 0.9$ 
  & $40.3 \pm 0.6$ 
  & $40.3 \pm 0.8$ \\
  \hline
  $\mathcal G(1)$ 
  & $1.044 \pm 0.006$ 
  & $1.041 \pm 0.006$ 
  & $1.054\pm0.009$
  & $1.044 \pm 0.006$
  & - \\
  \hline
  $\mathcal F(1)$ 
  & $0.900\pm0.009$
  & $0.895\pm0.011$
  & $0.904 \pm 0.012$
  & $0.895\pm0.011$
  & - \\
  \hline\hline
  \end{tabular} 
  }
  \caption{   
  Comparison of the $|V_{cb}|$ determinations along with the normalization factors $\mathcal F(1)$ and $\mathcal G(1)$. 
  In our work, these factors are simultaneously produced by the fit analysis. 
  }
  \label{Tab:comparisonVcb}
  \end{center}
\end{table}

We also provide a fit result for $\mathcal G(w)$ and $\mathcal F(w)$ comparing them with those in the CLN parameterization. 
The traditional form of $\mathcal G(w)$ is expanded by $z$, with the coefficients by means of the slope parameter $\rho^2$, and with the assumption estimated by UB as in Ref.~\cite{Caprini:1997mu}. 
In our study, we can directly produce the coefficients in $z$ expansion, defined as 
\begin{align}
 \label{Eq:normalizationG}
 \mathcal G(w) \equiv \mathcal G(1) \sum_{n=0}^3 g_n z^n \,,  
\end{align}
with $g_0=1$. 
Our result is then 
\begin{align}
   & g_1 = -7.77 \pm 0.43\,,& &g_2 = 24.9 \pm 5.3\,,& &g_3 = -38.6 \pm 33.0 \,,& 
\end{align}
and $\mathcal G(1) = 1.041 \pm 0.006$ for SM $(3/2/1)$, where the correlation matrix is put in Appendix~\ref{Ap:FitResult}. 
This is compared with the CLN form 
\begin{align}
 & g_1 = -8 \rho^2\,,& &g_2 = 51\rho^2-10\,,& &g_3 = -252\rho^2+84 \,,& 
\end{align}
for $\rho^2 = 1.131 \pm 0.033$~\cite{Amhis:2019ckw}. 
One can see that our result has $\sim 5$ times larger uncertainties in the coefficients. 
This is mainly due to inclusion of larger number of the parameters to be fitted.  
Thus, our result is rather conservative than the simple approximation of CLN as expected. 
Also, keep the discussion around Eq.~\eqref{Eq:expansion} in mind when $\mathcal G(w)^2$ is calculated for the evaluation of the decay rate.

The CLN form for $\mathcal F(w)$ is constructed with $h_{A_1}(w)$, $R_1(w)$, and $R_2(w)$.
As already explained, its CLN approximation is not appropriate for analyses with recent precise data. 
Instead, we provide the $z$ expanded $\mathcal F(w)$ squared such as 
\begin{align}
 \label{Eq:normalizationF}
 \mathcal F(w)^2 \equiv \mathcal F(1)^2 \sum_{n=0}^3 f^2_n z^n \,, 
\end{align}
with 
\begin{align}
  \mathcal F(1) = 0.895 \pm 0.011\,,\!\!&  & f_1^2 = -13.0 \pm 0.8 \,,\!\!& &f_2^2 = 55.9  \pm 18.0 \,,\!\!& &f_3^2 = -762 \pm 227 \,.
\end{align}

\subsubsection{NP scenarios}
We have seen the fit results including the NP contributions in Table~\ref{Tab:fitresult}. 
In the SM + $T$ scenarios, our fit result indicates that the $T$ contribution is constrained as $|C_T| < 0.025$ at $95\%$ confidence level for the case of $(3/2/1)$, which means zero-consistent, and hence omitted from the table. 
On the other hand, $|C_T| = 0.02 \pm 0.01$ is obtained for $(2/1/0)$ as seen from the table, which implies that the best fit point favors non-zero $T$ contribution although the uncertainty is still large. 
For both cases, the HQET parameters and $|V_{cb}|$ are then all consistent with those in the SM scenarios. 
This could be very interesting since the HQET parameterization model affects the fit result of the NP effect, and also the fit analysis has the NP sensitivity at the level of $\mathcal O (\%)$ of the SM value, $2\sqrt 2 G_F V_{cb}$.

The SM + $V_2$ scenarios also have the non-zero preferred value with the large uncertainty, $C_{V_2} = 0.05 \pm 0.01$ for $(3/2/1)$ and $C_{V_2} = 0.02 \pm 0.01$ for $(2/1/0)$. 
In addition, both cases give larger $|V_{cb}|$ than those in the SM, which would be interesting as it is different from the case for $\text{SM}+T$. 
Indeed, these changes improve the fit to the branching ratios. 
We obtain 
\begin{align}
 \mathcal B (\bar B^0 \to D^+ \ell^-\bar\nu)_{\text{SM}+V_2} & = \Big[  (2.29 \pm 0.06) \% \,; ~ (2.33 \pm 0.06) \%  \Big] \,, \\ 
 \mathcal B (\bar B^0 \to D^{*+} \ell^- \bar\nu)_{\text{SM}+V_2} & = \Big[  (5.05 \pm 0.05) \% \,; ~ (5.03 \pm 0.05) \%  \Big] \,, 
\end{align}
for $\text{SM}+V_2$ $\big[ (2/1/0);~ (3/2/1) \big]$ from the fit result. 
Thus we can see that the branching ratios are in perfect agreements with the experimental measurements.

The SM + $V_2$ + $T$ scenarios provide us with the most general model-independent fits. 
The $(3/2/1)$ case however results in the same as that of SM + $V_2$, inherited from the zero-consistent SM + $T$ result for $C_T$, and hence omitted from the table. 
The $(2/1/0)$ case finds non-zero best fit points for $C_{V_2}$ and $C_T$ by slightly changing the HQET parameters from the other scenarios. 
In this case, we obtain $\mathcal B (\bar B^0 \to D^+ \ell^-\bar\nu)_{\text{SM}+V_2+T}  = (2.29 \pm 0.06) \%$ and $\mathcal B (\bar B^0 \to D^{*+} \ell^- \bar\nu)_{\text{SM}+V_2+T} = (5.02 \pm 0.05) \%$.

The SM+NP fit improvements $\Delta \text{IC}$ for $(2/1/0)$ are a bit lower than the SM fit because of the additional degree, $C_X$. 
On the other hand, SM + $V_2$ $(3/2/1)$ has a clear improvement compared with SM $(3/2/1)$.

\begin{figure}[t!]
\begin{center}
\includegraphics[viewport=0 0 360 354, width=14em]{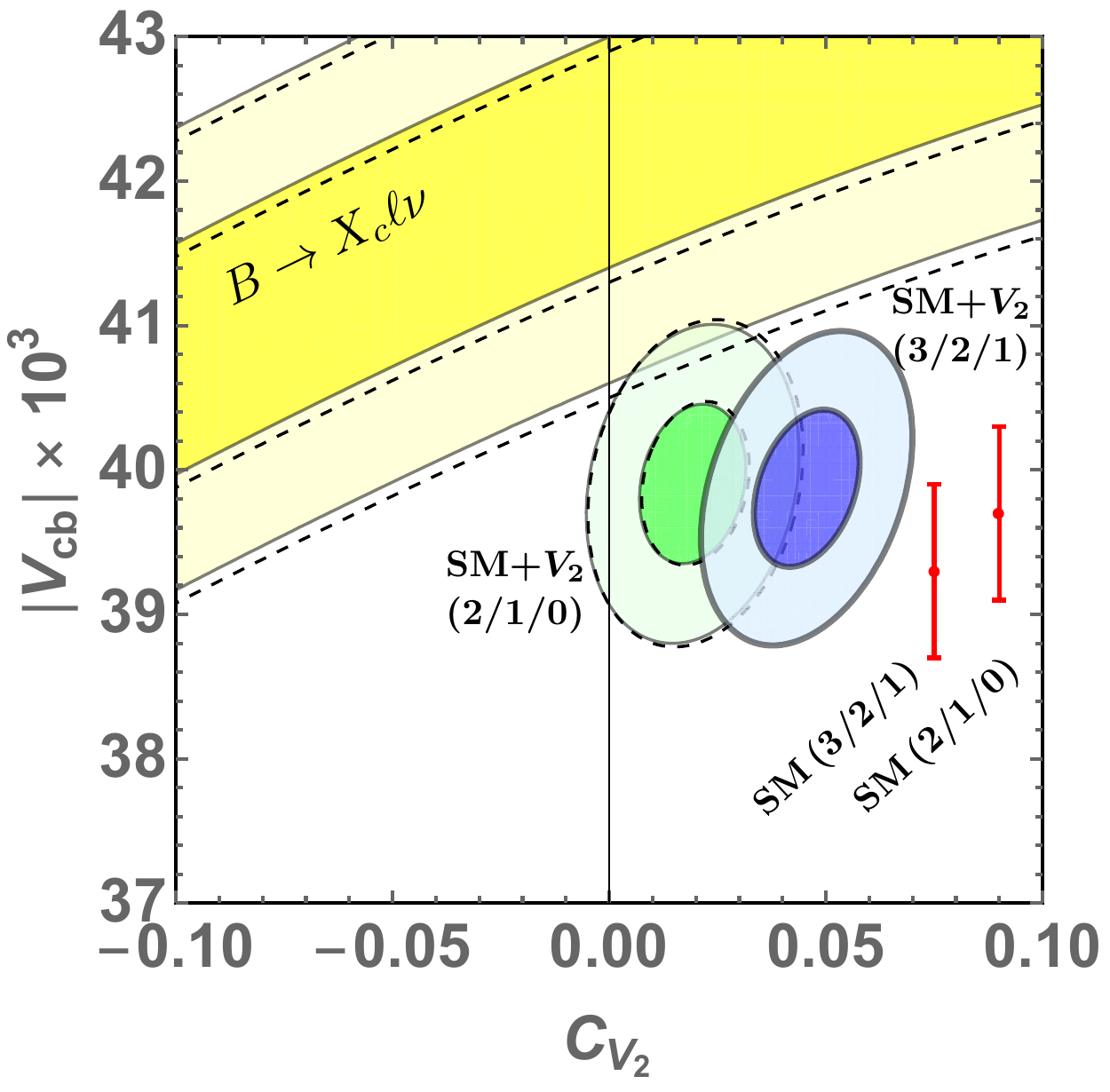} \quad\quad
\includegraphics[viewport=0 0 360 354, width=14em]{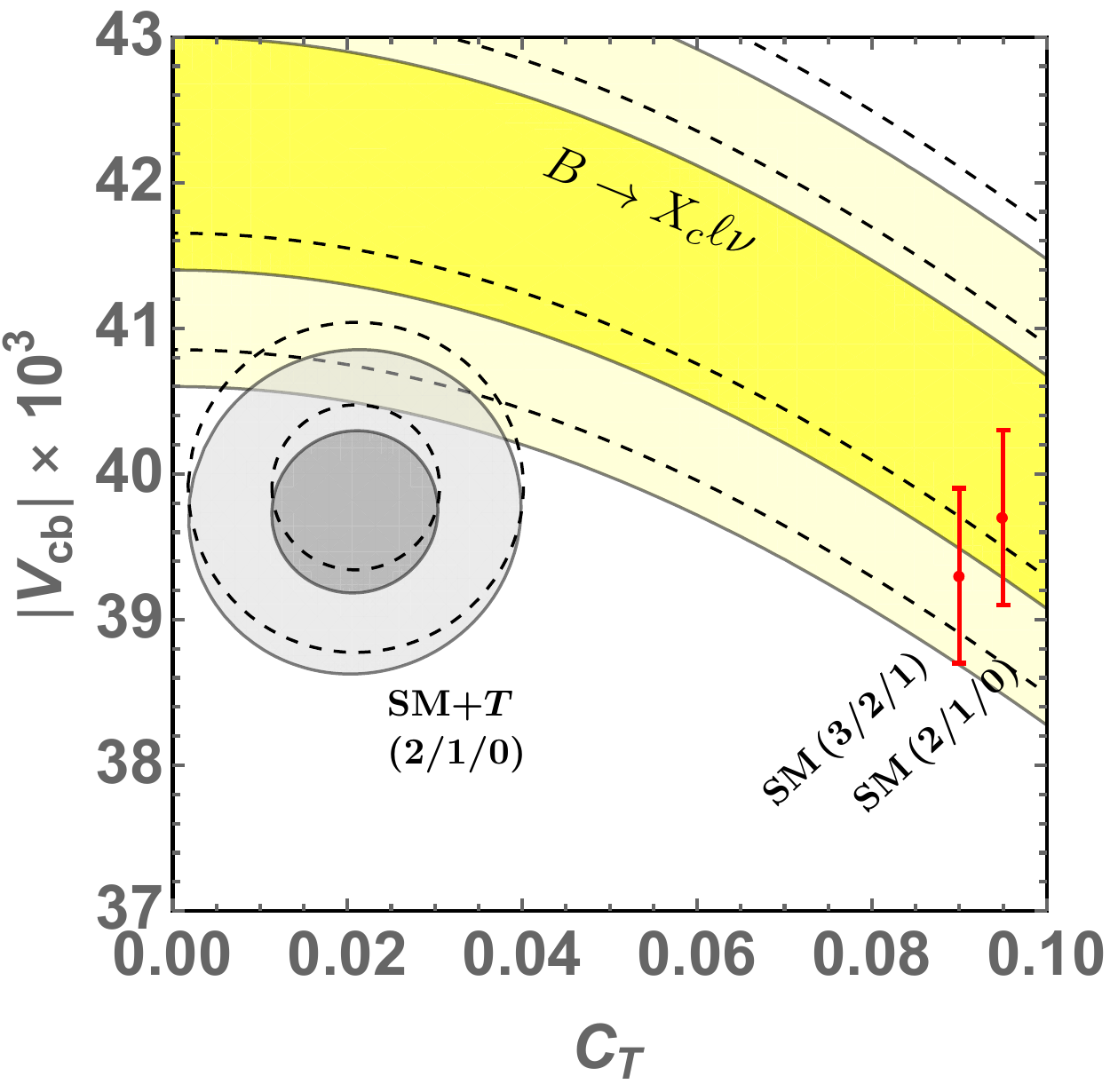} \\
\includegraphics[viewport=0 0 360 326, width=14.8em]{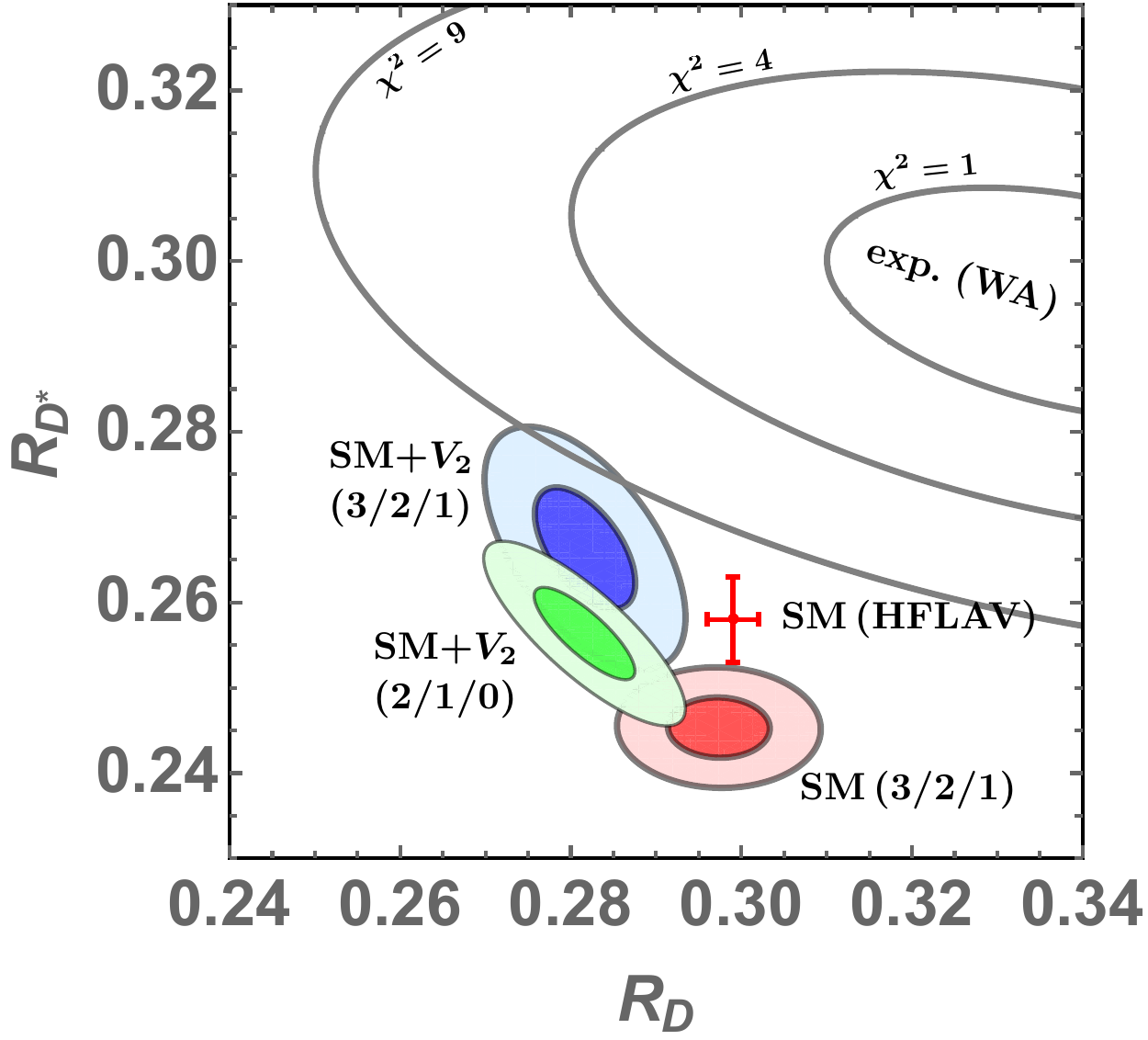}
\caption{
[Top]~preferred regions of $C_X$ and $|V_{cb}|$ in the SM + NP scenarios. 
The regions in blue, green, and gray are favored in the $\text{SM}+V_2 (3/2/1)$, $\text{SM}+V_2 (2/1/0)$, and $\text{SM}+T (2/1/0)$ scenarios, respectively. 
The yellow band indicates the allowed region from the inclusive mode. 
The contour lines correspond to $\Delta \chi^2 = 1, 4$. 
The red bars are the SM results for $|V_{cb}|$. 
The dashed lines are the case for SM + $V_2$ + $T$ by fixing another NP contribution to be $C_{T} = 0.05$ and $C_{V_2} = 0.02$ in the left and right panels, respectively.
[Bottom]~contour plot for predictions on $R_D$ and $R_{D^*}$ in the $\text{SM} (+ \text{NP})$ scenarios, 
where the regions for $\text{SM} (3/2/1)$, $\text{SM}+V_2 (3/2/1)$, and $\text{SM}+V_2 (2/1/0)$ are shown in red, blue, and green, respectively. 
The combined experimental result (gray solid curves that correspond to $\Delta\chi^2=1,4,9$) and the SM prediction in the literature (red bar) are taken from Ref.~\cite{Amhis:2019ckw}. 
} 
\label{Fig:VcbCT}
\end{center}
\end{figure}

In Fig.~\ref{Fig:VcbCT} (top), we show preferred regions of $|V_{cb}|$ and $C_X$ in the $\text{SM} + \text{NP}$ scenarios, 
where the regions in blue, green, and gray are favored in the $\text{SM}+V_2 (3/2/1)$, $\text{SM}+V_2 (2/1/0)$, and $\text{SM}+T (2/1/0)$ scenarios, respectively. 
We also include the allowed region from the inclusive process for $\text{SM} + \text{NP}$ as depicted in the yellow region. 
It can be derived with the use of Refs.~\cite{Jung:2018lfu,Crivellin:2009sd,Crivellin:2014zpa}, in which discrepancy of the $|V_{cb}|$ determination among the exclusive and inclusive processes has been investigated. 
Then, it is found that our fit result loosens the deviation in the $\text{SM}+V_2$ and $\text{SM}+T$ scenarios, but it is still not in a sufficient agreement. 
Lastly, one can see that the $\text{SM}+V_2+T$ scenario, displayed with the dashed lines, has no impact on this issue.

A corresponding LHC bound on $C_{X}$ is naively obtained by the following discussion. 
These days LHC constraints on NP effects have been getting severer. 
In Ref.~\cite{Greljo:2018tzh}, the authors have shown that $\tau$ searches with high $p_T$ at 36fb$^{-1}$~\cite{Aaboud:2018vgh,Sirunyan:2018lbg} give an upper limit on the WCs for the $b\to c\tau \nu$ current. 
(See, also Refs.\cite{Altmannshofer:2017poe,Iguro:2017ysu,Abdullah:2018ets,Iguro:2018fni,Baker:2019sli} in the context of NP interpretations of the $R_\Dgen$ anomaly.) 
Similarly, $e$ and $\mu$ searches with high $p_T$ at 139fb$^{-1}$~\cite{Aad:2019wvl} give an upper limit on $C_X$ defined as in Eq.~\eqref{Eq:effH}. 
Comparing those experimental constraints in looking at a tail of the $m_T$ plane ($\sim 1.4\text{TeV}$), we have the naive estimate of the upper bound as $|C_{V_2}| \lesssim 0.1$ and $|C_{T}| \lesssim 0.05$. 
Therefore, our fit result of $C_{X} \sim \mathcal O(0.01)$ is in the region of interest also for the LHC search. 
A further study of the LHC bound in higher $p_T$ ranges is work in progress.

\subsubsection{Observables for $\bar B \to \Dgen \tau\bar\nu$}
With the CLN parameterization, SM predictions and/or NP investigations have been provided with respect to $\bar B \to \Dgen \tau\bar\nu$ in the literature, ({\it e.g.}, see Refs.~\cite{Iguro:2018vqb,Blanke:2018yud} for recent works), 
since the experimental results have shown significant deviations from the SM predictions in the measurements of the ratios: $R_D = 0.340 \pm 0.027 \pm 0.013$ and $R_{D^*} = 0.295 \pm 0.011 \pm 0.008$ (combined average in Ref.~\cite{Amhis:2019ckw}). 
In recent years, challenging measurements of the $\tau$ and $D^*$ polarizations in $\bar B \to \Dst \tau\bar\nu$ have also been reported as $P_\tau^{D^*} = -0.38 \pm 0.51 \,^{+0.21}_{-0.16}$~\cite{Hirose:2016wfn} and $F_L^{D^*} = 0.60 \pm 0.08 \pm 0.04$~\cite{Abdesselam:2019wbt}.

Here, we also investigate these $\bar B \to \Dgen \tau\bar\nu$ observables with the use of our fit results in the $\text{SM} (+\text{NP})$ scenarios of the HQET parameterization. 
Note that, in this work, we mainly consider NP contributions only to the $(e ,\mu)$ modes.
Namely, the denominators of the ratios $R_\Dgen$ are only affected by the NP contribution. 
In this sense, our NP investigation has a different view from numerous previous studies for the $R_\Dgen$ anomaly, 
{\it e.g.}, see Refs.~\cite{Murgui:2019czp,Jaiswal:2020wer,Cheung:2020sbq} for the case of the HQET parameterization.
Also note that we take the proper $\epsilon_{a,b,c}$ expansion for these observables.

\begin{table}[t]
\renewcommand{\arraystretch}{1.3}
  \begin{center}
  \scalebox{0.85}{
  \begin{tabular}{lccccc}
  \hline\hline
  & $R_D$ & $R_{D^*}$ & $P_\tau^D$ & $P_\tau^{D^*}$ & $F_L^{D^*}$ \\
  \hline
  SM ($2/1/0$) & $0.289 \pm 0.004$ & $0.248 \pm 0.001$ & $0.331 \pm 0.004$ & $-0.496 \pm 0.007$ & $0.464 \pm 0.003$ \\ 
  \hline
  SM ($3/2/1$) & $0.297 \pm 0.006$ & $0.245 \pm 0.004$ & $0.326 \pm 0.003$ & $-0.503 \pm 0.020$ & $0.460 \pm 0.008$ \\
  \hline
  SM (HFLAV~\cite{Amhis:2019ckw}) & $0.299 \pm 0.003$ & $0.258 \pm 0.005$ & - & - & - \\
  \hline
  SM (Ref.\,\cite{Bordone:2019vic}) & $0.297 \pm 0.003$ & $0.250 \pm 0.003$ & $0.321 \pm 0.003$  & $-0.496 \pm 0.015$ & $0.464 \pm 0.010$ \\
  \hline\hline
  $\text{SM}+V_2$ ($2/1/0$) & $0.282 \pm 0.006$ & $0.256 \pm 0.005$ & $0.332 \pm 0.004$ & $-0.499 \pm 0.007$ & $0.465 \pm 0.003$ \\
  LFU case & $0.292 \pm 0.004$ & $0.247 \pm 0.001$ & $0.332 \pm 0.004$ & $-0.499 \pm 0.007$ & $0.463 \pm 0.003$ \\
  \hline
  $\text{SM}+V_2$ ($3/2/1$) & $0.282 \pm 0.006$ & $0.266 \pm 0.007$ & $0.329 \pm 0.003$ & $-0.506 \pm 0.020$ & $0.464 \pm 0.008$ \\
  LFU case & $0.309 \pm 0.006$ & $0.244 \pm 0.004$ & $0.329 \pm 0.003$ & $-0.507 \pm 0.020$ & $0.460 \pm 0.008$ \\
  \hline\hline
  \end{tabular} 
  }
  \caption{   
  Predictions of the $\bar B \to \Dgen \tau\bar\nu$ observables.
  }
  \label{Tab:tauresult}
  \end{center}
\end{table}

In Table~\ref{Tab:tauresult}, we list our predictions on the $\bar B \to \Dgen \tau\bar\nu$ observables in the present models, along with those from Refs.~\cite{Amhis:2019ckw,Bordone:2019vic}. 
Our analysis shows that the $\text{SM} (2/1/0)$ predicts smaller values for both $R_D$ and $R_\Dst$ than those of the HFLAV report. 
On the other hand, the $\text{SM} (3/2/1)$ has the consistent value for $R_D$ while smaller for $R_\Dst$. 
This is a similar behavior with that obtained in Ref.~\cite{Bordone:2019vic}. 
Then, it is also found that the polarizations for $(3/2/1)$ are consistent with the reference. 
We obtain the same results for the cases of $\text{SM}+T$.

The $\text{SM}+V_2$ models give $R_\Dgen$ different from the SM predictions, which could be NP signals at the Belle~II experiment with large statistics. 
To be precise, both cases point to the $R_D$ and $R_{D^*}$ values smaller and larger than the SM predictions, respectively. 
In particular, it is interesting that the $R_\Dst$ deviation from the measurement becomes smaller for $\text{SM}+V_2 (3/2/1)$. 
This is a key feature for this model. 
However, we have to remark that the experimental measurement for $R_\Dgen$ is analyzed by means of both $\tau$ and $(e,\mu)$ distributions subtracted from background. 
The present measurement is then done with the assumption that the $(e,\mu)$ modes obey the SM.  
Thus, in the presence of NP in the $(e,\mu)$ modes, re-analysis is needed by taking the NP effect. 
Although such a NP effect of $\mathcal O(\%)$ is negligible for the present analysis, it could become significant as larger number of events are accumulated at the Belle~II experiment.

Predicted values of the tau observables in the $\text{SM}+V_2+T$ scenarios are all equivalent to those in $\text{SM}+V_2$ within the present uncertainties.

Finally, we show a summary plot for the predictions on $R_D$ and $R_{D^*}$ in the $\text{SM} (+ \text{NP})$ scenarios in Fig.~\ref{Fig:VcbCT} (bottom), 
where the allowed regions for $\text{SM} (3/2/1)$, $\text{SM}+V_2 (3/2/1)$, and $\text{SM}+V_2 (2/1/0)$ are shown in red, blue, and green, respectively. 
The combined experimental result (gray curves) and the referred SM prediction (red bar) are taken from Ref.~\cite{Amhis:2019ckw}.

One might be interested in the case where the NP contribution is universal for all the leptons (LFU), such that $C_X^e = C_X^\mu = C_X^\tau$.
We also show predictions for such a case in the table. 
For $\text{SM}+V_2$, $(1 \pm C_{V_2})^2$ enters as the overall factor both in the numerator and denominator of $R_D/R_{D^*}$, 
and thus the values should be equivalent to the SM ones if using the same inputs for the HQET parameters. 
(To be precise, $R_{D^*}$ is affected by $C_{V_2}$ beyond the overall factor, but such an effect is quite small.)
Our $V_2$ LFU results are not the same as our SM values due to the differences in the fitted HQET parameters which can be seen in Table~\ref{Tab:fitresult}. 
In particular, the $(3/2/1)$ model predicts larger $R_D$ than the other cases and scenarios, which could be interesting. 
We also found that $T$ LFU results are all similar to the corresponding SM values. 
In any event, however, our results show a fact that the present $R_\Dgen$ anomaly is not resolved by the LFU type NP contribution. 
Therefore, some large $C_X^\tau$ violating LFU is necessary.

\subsubsection{Theoretical uncertainty}
\label{Sec:thuncertainty}
\begin{figure}[t!]
\begin{center}
\includegraphics[viewport=0 0 715 321, width=24em]{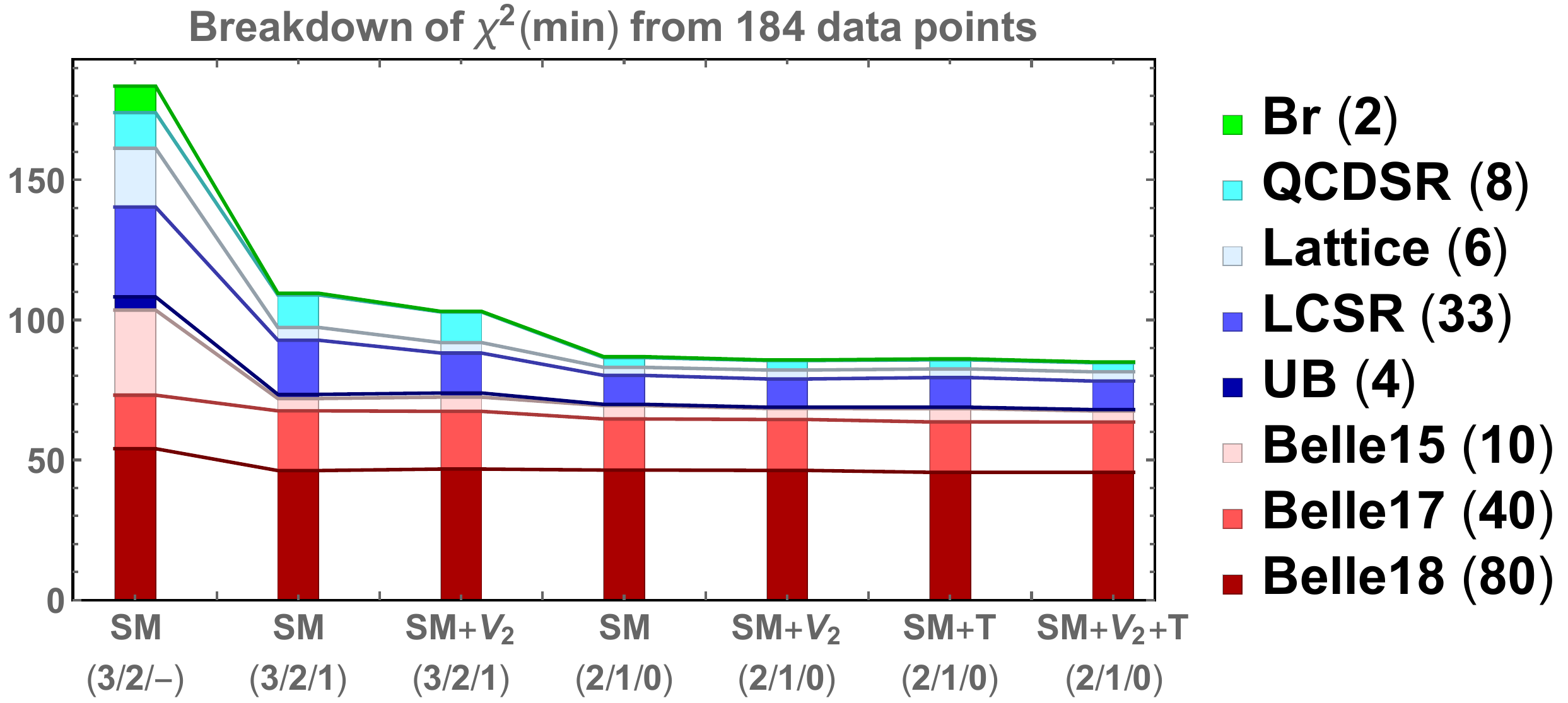} 
\caption{
The breakdown of the $\chi^2$ deviations at our best fit results from the 184 data points. 
} 
\label{Fig:ChiSqBreakdown}
\end{center}
\end{figure}

For the present analyses so far, we have treated the theory constraints as being normally distributed for simplicity and in order to obtain  the applicable outputs. 
Thanks to it, we can display the breakdown of the $\chi^2$ deviations for our fit results as in Fig.~\ref{Fig:ChiSqBreakdown}.

The UBs are taken as the Gaussian distribution assuming the central value as zero while the standard deviation as the calculated upper limit given in Eqs.~\eqref{Eq:UB1}--\eqref{Eq:UB4}. 
We have checked the breakdown of the $\chi^2$ deviation from the UBs for all the present models considered in our analysis and then have confirmed that those for the UBs are all within $1\sigma$.

As for the bounds from QCDSR, we have derived the constraints of Eqs.~\eqref{Eq:QCDSR1}--\eqref{Eq:QCDSR3} and again taken as normal distributions. 
Our fit results, however, show that some of the NLO parameters are deviated from these constraints as seen in Table~\ref{Tab:fitresult}. 
In particular, our MCMC run finds the best fit point of $\IWci{1}{3} \sim \mathcal O(-0.01)$ that has a large deviation from the QCDSR constraint $\sim \mathcal O(+0.01)$. 
Indeed, the $\chi^2$ breakdown for the QCDSR constraints is $\chi^2_\text{QCDSR} \sim 10 (4)$ for the SM $(3/2/1)$ (SM $(2/1/0)$). 
To see its effect, we test a fit analysis where possible ranges for the NLO parameters, $\IWci{n}{2,3}$ and $\IWet{n}$, are restricted as in Eqs.~\eqref{Eq:QCDSR1}--\eqref{Eq:QCDSR3}. 
Then we find that the outputs of $|V_{cb}|$ and the LO parameters $\IWxi{n}$ are not much affected while those of the NNLO parameters $\IWel{n}{i}$ are shifted, compared with the results obtained in Table~\ref{Tab:fitresult}. 
In this case, however, the NLO parameter fits have bad convergences and their distributions are far away from the normal distributions. 
Also, we have checked that the observables such as the branching ratios are all consistent.
In this sense, we can say that our main conclusion is not affected by this issue.

\section{Summary}
\label{Sec:conclusion}
We have investigated the semi-leptonic decays of $\bar B \to \Dgen \ell\bar\nu$ in terms of the HQET parameterization for the form factors, 
with the heavy quark expansion up to $\mathcal O(1/m_c^2)$, and beyond the simple approximation considered in the original CLN parameterization. 
It is given with the $z = (\sqrt{w+1}-\sqrt2) / (\sqrt{w+1}+\sqrt2)$ expanded form, and then the highest order for the expansion is in principle arbitrary. 
In our work, we have followed the models from Ref.~\cite{Bordone:2019vic} denoted as $(2/1/0)$ and $(3/2/1)$ for the $z$ expansions in the (leading/sub-leading/subsub-leading) IW functions.

The analysis with this setup was first given in Ref.~\cite{Bordone:2019vic}, and then we have extended it to the comprehensive analyses including 
(i) simultaneous fit of $|V_{cb}|$ and the HQET parameters to the available experimental full distribution data and the theory constraints, and 
(ii) NP contributions of the $V_2$ and $T$ types, such as $(\overline{c} \gamma^\mu P_Rb)(\overline{\ell} \gamma_\mu P_L \nu_{\ell})$ and $(\overline{c}  \sigma^{\mu\nu}P_Lb)(\overline{\ell} \sigma_{\mu\nu} P_L \nu_{\ell})$, to the decay distributions and rates. 
For this purpose, we have performed the Bayesian fit analyses by using {\tt Stan} program, a state-of-the-art public platform for statistical computation, in which MCMC runs with various algorithms are possible.

Then it has been shown that our $|V_{cb}|$ fit results for the SM $(2/1/0)$ and $(3/2/1)$ scenarios are both close to the PDG combined average, $(39.5 \pm 0.9) \times 10^{-3}$, from the exclusive mode: 
\begin{align}
 |V_{cb}|_{(2/1/0)} & = (39.7 \pm 0.6) \times 10^{-3} \,, \\
 |V_{cb}|_{(3/2/1)} & = (39.3 \pm 0.6) \times 10^{-3} \,, 
\end{align}
as also summarized in Table~\ref{Tab:comparisonVcb}.
This implies that the deviation from the inclusive mode still holds.
We have also found that the fit to the $w$ distribution data with the theory constraints ({\bf\boldmath $w+$theory}) reproduce the larger $|V_{cb}|$ value completely consistent with that reported in Ref.~\cite{Bordone:2019vic}. 
This could imply significance of the angular distribution data for $\bar B \to \Dst \ell\bar\nu$. 
Besides, we have evaluated {\it Information Criterion} to see how the inclusion of the $\mathcal O(1/m_c^2)$ parameters improve the fit. 
Then we see that the 23 HQET parameters of $(3/2/1)$ are surplus while 13 of $(2/1/0)$ are sufficient for the statistical modeling to explain the present available data points.

The $\text{SM} + \text{NP}$ scenarios have been studied with the same manner. 
At first, we have confirmed that $\text{SM} + T (3/2/1)$ is  constrained as $|C_T| < 0.025$ at $95\%$ confidence level and the best fit point is zero-consistent. 
On the other hand, it has turned out that $\text{SM} + T (2/1/0)$ is allowed to have non-zero contribution, $|C_T| = 0.02 \pm 0.01$, to the processes.
This could be very interesting since the HQET parameterization model affects the fit result of the NP effect. 
Furthermore, a significant point is that the fit analysis has the NP sensitivity at the level of $\mathcal O (\%)$.

Then, we have also obtained non-zero preferred values in the $\text{SM} + V_2$ scenarios as $C_{V_2} = 0.05 \pm 0.01$ for $(3/2/1)$ and  $C_{V_2} = 0.02 \pm 0.01$ for $(2/1/0)$. 
{\it Information Criterion} also suggests that $\text{SM} + V_2$ is favored at the same level with the SM scenarios. 
In addition, both cases give larger $|V_{cb}|$ than those in the SM, but they are still not in a sufficient agreement with the $|V_{cb}|$ determination from the inclusive process. 
In turn, we have considered $\text{SM} + V_2 + T$ for the model-independent fit. 
We then found that $C_T$ is zero-consistent and the result is the same as that of $\text{SM} + V_2$ for the $(3/2/1)$ scenario. 
The $(2/1/0)$ scenario points to non-zero $C_{V_2}$ and $C_T$ with the same $|V_{cb}|$ as that of $\text{SM} + V_2$.
This is summarized in Fig.~\ref{Fig:VcbCT} (top). 
The applicable LHC bound is naively given as $|C_{V_2}| \lesssim 0.1$ and $|C_{T}| \lesssim 0.05$ estimated from the $m_T$ plane at $\sim 1.4\text{TeV}$ and thus a further LHC search would be interesting.

Finally, we have produced our predictions on the $\bar B \to \Dgen \tau\bar\nu$ observables in the present models. 
It is summarized in Table~\ref{Tab:tauresult} and Fig.~\ref{Fig:VcbCT} (bottom). 
Our prediction in $\text{SM} (2/1/0)$ has smaller values for both $R_D$ and $R_\Dst$ comapred with those in the HFLAV report. 
On the other hand, $\text{SM} (3/2/1)$ predicts the consistent value for $R_D$ while smaller for $R_\Dst$. 
In the $\text{SM} + V_2$ scenarios, NP only contributes to the light-lepton modes and then it results in the $R_D$ and $R_{D^*}$ values smaller and larger than the SM predictions, respectively. 
It is also seen that the $R_\Dst$ deviation from the experimental measurement becomes milder ($\sim1.9\sigma$) than the one in the SM. 
This is a key feature for this model derived from our fit analysis.  
Quantitatively, we have $4.24\sigma$ and $3.66\sigma$ significances for the $R_\Dgen$ deviation in SM $(2/1/0)$ and $(3/2/1)$, respectively. 
This is compared with $3.03\sigma$ in SM (HFLAV) assuming no correlation. 
As for $\text{SM} + V_2$, we see $3.76\sigma$ $(2/1/0)$ and $2.96\sigma$ $(3/2/1)$.  
For a specific interest, we have checked the case where the NP contribution is lepton-flavor universal, $C_X^e = C_X^\mu = C_X^\tau$ for the tau observables.
Then we saw that $\text{SM} + V_2 (3/2/1)$ with LFU predicts larger $R_D$ value ($= 0.309 \pm 0.06$) than any other models and scenarios, which could be interesting.

For the present analyses, we have treated the theory constraints as being normally distributed for simplicity and in order to obtain the applicable outputs. 
A further practical treatment on the theoretical uncertainties could be possible, for instance, when we implement this work in the public {\tt HEPfit} package~\cite{deBlas:2019okz}. 
We leave it for our future work.

We conclude from this work that the available full distribution data of $\bar B \to \Dgen \ell\bar\nu$ has potential to fit a large number of the parameters in the HQET parameterization together with $|V_{cb}|$, and a further improvement is expected at the Belle~II experiment. 
The fit analysis also has the NP sensitivity with the $\mathcal O(\%)$ level of the SM contribution, and then it could be examined with the $\bar B \to \Dgen \tau\bar\nu$ observables in future.
Interesting directions of future work are, for example, CP violation~\cite{Bhattacharya:2019olg} and QED corrections~\cite{deBoer:2018ipi} in the $\bar B \to \Dgen \ell\bar\nu$ distributions with respect to the HQET parameterization.

\section*{Acknowledgements}
We are grateful to Martin Jung for useful comments on the HQET parameterization. 
We thank Minoru Tanaka for discussion about QCDSR. 
RW thanks Vincent Picaud for helpful suggestions on the usage of {\tt MathematicaStan}. 
RW also thanks Marco Ciuchini for discussion on MCMC. 
SI is grateful to Kazuhiro Tobe for discussion about various aspects of this work. 
SI also thanks Kodai Matsuoka and Noritsugu Tsuzuki for useful comments on the experimental measurements of $\bar B \to \Dgen \ell\bar\nu$ and $\bar B \to \Dgen \tau\bar\nu$. 
We also thank Nagoya University Theoretical Elementary Particle Physics Laboratory for providing computational resources. 
For version 4: we thank Hongkai Liu for pointing out our typos in the previous version, and Paolo Gambino for discussion about UB. 
The work of SI is supported by the Japan Society for the Promotion of Science (JSPS) Research Fellowships for Young Scientists, No. 19J10980 and Core to Core Program, No.  JPJSCCA20200002.

\appendix
\section*{Appendix}
\section{$\alpha_s$ and $1/m_Q$ Corrections}
\label{Ap:Corrections}
Here we list functions for the $\alpha_s$ and $1/m_{b,c}$ corrections. 
We have followed the analytic result from Ref.~\cite{Bernlochner:2017jka}. 
The $\alpha_s$ corrections, $\dhh_{X,\alpha_s}$, are given as 
\begin{align}
 \dhh_{+,\alpha_s} = 
 {1 \over 6\zcb^2 (w - \wcb)^2} 
 &\, \left[ 4\zcb^2 (w-\wcb)^2 \Omega_w (w) + (w+1) \cred{\big(} -1 +(w+w^2+2\wcb) (\zcb^2+1) \zcb \right. \notag \\
 &\,\hspace{1em} + 2(w^2-w(2+3\wcb)+\wcb)\zcb^2 - \zcb^4 \cred{\big)} r_w (w) \notag \\
 &\,\hspace{1em} +(\wcb-w) \cblu{\big(} 1+w+w(10+\zcb) \zcb +(-2-12\wcb+\zcb)\zcb \cblu{\big)} \zcb \notag \\
 &\,\hspace{1em}  \left. -  (1+\wcb-w-w^2) (1-\zcb^2)\zcb \log \zcb \right] + V(\mu) \,, 
\end{align}
\begin{align}
 \dhh_{-,\alpha_s} = 
 {1+w \over 6\zcb^2 (w - \wcb)^2} 
 &\, \left[- \big(1 +(1-2w-w^2)\zcb +\zcb^2\big)(1-\zcb^2)r_w (w) -(w-\wcb) (1-\zcb^2)\zcb \right. \notag \\
 &\, \left. -\cred{\big(}2(1+\zcb+\zcb^2) -w(1+4\zcb+\zcb^2)\cred{\big)}\zcb \log \zcb \right] \,, 
\end{align}
\begin{align}
 \dhh_{V,\alpha_s} = 
 {1 \over 6\zcb (w - \wcb)} 
 &\, \left[ 4\zcb (w-\wcb) \Omega_w (w) + 2(w+1)((3w-1)\zcb -\zcb^2-1) r_w (w) \right. \notag \\
 &\, \left. -12\zcb (w-\wcb) -(\zcb^2-1) \log \zcb \right] + V(\mu) \,, 
\end{align}
\begin{align}
 \dhh_{A_1,\alpha_s} = 
 {1 \over 6\zcb (w - \wcb)} 
 &\, \left[ 4\zcb (w-\wcb) \Omega_w (w) + 2(w-1)((3w+1)\zcb -\zcb^2-1) r_w (w) \right. \notag \\
 &\, \left. -12\zcb (w-\wcb) -(\zcb^2-1) \log \zcb \right] + V(\mu) \,, 
\end{align}
\begin{align}
 \dhh_{A_2,\alpha_s} = 
 {\cred{-1} \over 6\zcb^2 (w - \wcb)^2} 
 &\, \left[ \left(2 + (2w^2-5w-1)\zcb + 2w(2w-1)\zcb^2 + (1-w)\zcb^3 \right) r_w (w) \right. \notag \\
 &\, \left. -2\zcb(\zcb+1)(w-\wcb) + (\zcb^2-(4w+2)\zcb +3 +2w) \zcb \log \zcb \right] \,, \label{eq:mod2}
\end{align}
\begin{align}
 \cred{\dhh_{A_3,\alpha_s}} = 
  \dhh_{A_1,\alpha_s} 
 & + \frac{1}{6\zcb (w - \wcb)^2}  \bigl{[} \,2 \zcb (\zcb +1) (\wcb - w) \nonumber \\ 
 & + \big( 2\zcb^3 + \zcb^2 (2w^2-5w-1) + \zcb (4w^2-2w) - w +1 \big) r_w (w) \nonumber \\[0.5em]
 & -\big( \zcb^2 (2w+3) - \zcb (4w+2) +1 \big) \log \zcb \bigl{]} \label{eq:mod3}
 \end{align}
\begin{align}
 \dhh_{S,\alpha_s} = 
 {1 \over 3\zcb (w - \wcb)} 
 &\, \left[ 2\zcb (w-\wcb) \Omega_w (w) - (w-1)(\zcb+1)^2 r_w (w) \right. \notag \\
 &\, \left. + (\zcb^2-1) \log \zcb \right] + S(\mu) \,, 
\end{align}
\begin{align}
 \dhh_{P,\alpha_s} = 
 {1 \over 3\zcb (w - \wcb)} 
 &\, \left[ 2\zcb (w-\wcb) \Omega_w (w) - (w+1)(\zcb-1)^2 r_w (w) \right. \notag \\
 &\, \left. + (\zcb^2-1) \log \zcb \right] + S(\mu) \,, 
\end{align}
\begin{align}
 \dhh_{T,\alpha_s} = 
 {1 \over 3\zcb (w - \wcb)} 
 &\, \left[ 2\zcb (w-\wcb) \Omega_w (w) + \cred{\big(} 4\zcb w^2 -(1-\zcb)^2 w -(1+\zcb)^2 \cred{\big)} r_w (w) \right. \notag \\
 &\, \left. -6\zcb(w-\wcb) + (1-\zcb^2) \log \zcb \right] +T(\mu) \,, 
\end{align}
\begin{align}
 \dhh_{T_1,\alpha_s} = 
 {1 \over 3\zcb (w - \wcb)} 
 &\, \left[ 2\zcb (w-\wcb) \Omega_w (w) + 2\zcb (w^2-1) r_w (w) \right. \notag \\
 &\, \left. -6\zcb (w-\wcb) +(1-\zcb^2) \log \zcb \right] +T(\mu) \,, 
\end{align}
\begin{align}
 \dhh_{T_2,\alpha_s} = 
 {w+1 \over 3\zcb (w - \wcb)} 
 \left[ (1-\zcb^2) r_w (w) +2 \zcb \log \zcb \right] \,, 
\end{align}
\begin{align}
 \dhh_{T_3,\alpha_s} = 
 {1 \over 3\zcb (w - \wcb)} 
 \left[ (\zcb w-1) r_w (w) - \zcb \log \zcb \right] \,, 
\end{align}
with 
\begin{align}
 & \zcb = {m_c \over m_b} \,, \quad \wcb = {1\over2} \left( \zcb + \zcb^{-1} \right) \,, \quad w_\pm(w) = w \pm \sqrt{w^2-1} \,, \\
 & r_w (w) = {\log w_+ (w) \over \sqrt{w^2-1} }\,, \\
 & \Omega_w (w) = {w \over 2\sqrt{w^2-1}} \Big[2\text{Li}_2 (1-w_-(w)\zcb) - 2\text{Li}_2 (1-w_+(w)\zcb) \notag \\
 & \hspace{9em}+ \text{Li}_2 (1-w_+^2(w)) - \text{Li}_2 (1-w_-^2(w)) \Big] - w r_w(w) \log \zcb + 1 \,, 
\end{align}
where $\text{Li}_2(x) = \int_x^0 dt \log(1-t)/t$ is dilogarithmical function. 
The above results are obtained at the scale $\mu_{\sqrt{bc}} = \sqrt{m_bm_c}$, namely $V(\mu_{\sqrt{bc}})=S(\mu_{\sqrt{bc}})=T(\mu_{\sqrt{bc}})=0$. 
Otherwise, the scale factors are given as 
\begin{align}
 V(\mu) & = -{2\over3} \big( wr_w(w)-1 \big) \log {m_bm_c \over \mu^2} \,, \\
 S(\mu) & = -{1\over3} \big( 2wr_w(w)+1 \big) \log {m_bm_c \over \mu^2} \,, \\
 T(\mu) & = -{1\over3} \big( 2wr_w(w)-3 \big) \log {m_bm_c \over \mu^2} \,. 
\end{align}
Note that we set the scale as $\mu_b = 4.2\text{GeV}$ in our analysis.

The $1/m_{b,c}$ corrections involve four sub-leading IW functions, $\chi_{1\text{-}3}(w)$ and $\xi_3(w)$, one of which (usually $\chi_1$) can be absorbed into the definition of $\xi(w)$.  
For the form of $\dhh_{X,m_{b,c}}$, the sub-leading IW functions divided by $\xi(w)$ are defined as in Eq.~\eqref{Eq:subIW}. 
Following Ref.~\cite{Bernlochner:2017jka}, we can write $\dhh_{X,m_{b,c}}$ as 
\begin{align}
 \dhh_{+,m_b} = 
 \dhh_{+,m_c} = 
 \dhh_{T_1,m_b} = 
 -4(w-1) \hat\chi_2(w) + 12 \hat\chi_3(w) \,, 
\end{align}
\begin{align}
 \dhh_{-,m_b} = 
 -\dhh_{-,m_c} =
 \dhh_{T_2,m_b} =
 1 - 2\eta(w) \,, 
\end{align}
\begin{align}
 \dhh_{V,m_b} 
  = &~ \dhh_{A_3,m_b} = \dhh_{P,m_b} = \dhh_{T,m_b} = \dhh_{T,m_c}  \notag\\
  = &~ 1 - 2 \eta(w) -4(w-1) \hat\chi_2(w) + 12 \hat\chi_3(w) \,, 
\end{align}
\begin{align}
 \dhh_{V,m_c} = 
 & 1 - 4 \hat\chi_3(w) \,, 
\end{align}
\begin{align}
 \dhh_{A_1,m_b} 
 = &~ \dhh_{S,m_b} = \dhh_{S,m_c} \notag \\
 = &~ (w-1) \Big[ (w+1)^{-1} \big( 1-2\eta(w) \big) -4 \hat\chi_2(w) \Big] +12 \hat\chi_3(w)  \,, 
\end{align}
\begin{align}
 \dhh_{A_1,m_c} 
 = &\, (w-1) (w+1)^{-1} -4\hat\chi_3(w)  \,, 
\end{align}
\begin{align}
 \dhh_{A_2,m_b} = &\, \dhh_{T_3,m_b} = 0 \,, 
\end{align}
\begin{align}
 \dhh_{A_2,m_c} = 
 & -2(w+1)^{-1} \big( 1 + \eta(w) \big) + 4 \hat\chi_2(w)  \,, 
\end{align}
\begin{align}
 \dhh_{A_3,m_c} = 
 &\, 1 - 2 (w+1)^{-1} \big( 1 + \eta(w) \big) -4 \hat\chi_2(w) -4\hat\chi_3(w) \,, 
\end{align}
\begin{align}
 \dhh_{P,m_c} = 
 &\, - 1 + 2 \big( 1+\eta(w) \big) +4(w-1) \hat\chi_2(w) - 4 \hat\chi_3(w) \,, 
\end{align}
\begin{align}
 \dhh_{T_1,m_c} & = -4 \hat\chi_3(w) \,, 
\end{align}
\begin{align}
 \dhh_{T_2,m_c} & = -1 \,, 
\end{align}
\begin{align}
 \dhh_{T_3,m_c} & = (w+1)^{-1} \big( 1 + \eta(w) \big) + 2 \hat\chi_2(w) \,, 
\end{align}
where $\chi_1(w)$ is absorbed.

The $1/m_c^2$ corrections consist of six subsub-leading IW functions $\ell_{1\text{-}6}(w)$ in the absence of the $1/m_b^2$ and  $1/(m_bm_c)$  corrections~\cite{Falk:1992wt}. 
The expressions for $\dhh_{X,m_c^2}$ can be obtained from Ref.~\cite{Falk:1992wt} as 
\begin{align}
 \dhh_{+,m_c^2} & = \hat\ell_1 (w) \,, \\
 \dhh_{-,m_c^2} & = \hat\ell_4 (w) \,, \\
 \dhh_{V,m_c^2} & = \hat\ell_2 (w) - \hat\ell_5 (w) \,, \\
 \dhh_{A_1,m_c^2} & = \hat\ell_2 (w) - {w-1 \over w+1} \hat\ell_5 (w) \,, \\
 \dhh_{A_2,m_c^2} & = \hat\ell_3 (w) + \hat\ell_6 (w) \,, \\
 \dhh_{A_3,m_c^2} & = \hat\ell_2 (w) - \hat\ell_3 (w) - \hat\ell_5 (w) + \hat\ell_6 (w)  \,, 
\end{align}
\begin{align}
 \dhh_{S,m_c^2} & = \hat\ell_1 (w) - {w-1 \over w+1} \hat\ell_4 (w) \,, \\
 \dhh_{P,m_c^2} & = \hat\ell_2 (w) + (w-1) \hat\ell_3 (w) + \hat\ell_5 (w) - \cred{(w+1)} \hat\ell_6 (w) \,, \label{eq:mod4}
\end{align}
\begin{align}
 \dhh_{T,m_c^2} & = \hat\ell_1 (w) - \hat\ell_4 (w) \,, \\
 \dhh_{T_1,m_c^2} & = \hat\ell_2 (w) \,, \\
 \dhh_{T_2,m_c^2} & = \hat\ell_5 (w) \,, \\
 \dhh_{T_3,m_c^2} & = {1\over2} \big( \hat\ell_3 (w) -\hat\ell_6 (w) \big) \,, 
\end{align}
for $\hat\ell(w) = \ell(w) / \xi(w)$.

\section{Angular dependence}
\label{Ap:AngularFormula}
Here we derive the full angular distribution of Eqs.~\eqref{eq:fulldistribution} and \eqref{eq:angular}. 
In the SM, the squared decay amplitude of $\mathcal M$ for $B^0 \to \Dst^- \ell\bar\nu$, followed by $\Dst^- \to \bar D^0 \pi^-$, can be represented as 
\begin{align}
 |\mathcal M(q^2,\theta_\ell, \theta_V, \chi)|^2 = \left| \sum_{\lambda_{\Dst},\lambda'_{\Dst}} \mathcal S^{\lambda_{\Dst}} (\theta_\ell)\, \mathcal D^1_{\lambda_{\Dst},\,\lambda'_{\Dst}}(\chi)\, \mathcal T^{\lambda'_{\Dst}}(\theta_V) \right|^2 \,, 
\end{align}
where  
\begin{align}
 \mathcal S^{\lambda_{\Dst}} (q^2,\theta_\ell) = {G_F \over \sqrt{2}} V_{cb} \sum_{\lambda_W} H^{\lambda_{\Dst}}_{\lambda_W} L_{\lambda_W}^{\lambda_\ell=-1/2} \,, 
\end{align}
shows the usual helicity amplitude for $B^0 \to \Dst^- \ell\bar\nu$, which has been described in Ref.~\cite{Tanaka:2012nw},  
whereas $\mathcal T^{\lambda'_{\Dst}}(\theta_V)$ indicates the amplitude for $\Dst^- \to \bar D^0 \pi^-$ 
and $\mathcal D^1_{\lambda_{\Dst}\,\lambda'_{\Dst}}(\chi)$ is the Wigner rotation that connects two decay planes defined for $\theta_\ell$ ($\ell$-$\nu$ plane at $W$ rest frame) and $\theta_V$ ($D$-$\pi$ plane at $\Dst$ rest frame). 
Then the latter two can be obtained as 
\begin{align}
 \mathcal T^{0} = {N\over2} {3 \over \pi} \cos\theta_V \,, \quad \mathcal T^{\pm1}  = \mp {N\over2} {3 \over 2 \pi} \sin\theta_V \,, 
\end{align}
and 
\begin{align}
 \mathcal D^1_{0,0} = 1 \,, \quad  \mathcal D^1_{\pm1,\pm1} = e^{\pm i \chi} \,, \quad \text{others} = 0 \,, 
\end{align}
where the normalization factor $N$ is determined so that 
\begin{align}
 \int_{-1}^1 d\cos\theta_V\int_{-\pi}^{\pi} d\chi {d\Gamma_\text{full} \over dw\,d\cos\theta_\ell\,d\cos\theta_V\,d\chi} = {d\Gamma ({B^0 \to \Dst^- \ell\bar\nu}) \over dw\,d\cos\theta_\ell} \mathcal B ({\Dst^- \to \bar D^0 \pi^-}) \,, 
\end{align}
is satisfied. 
Following $L_{\lambda_W}^{\lambda_\ell=-1/2}$ by substituting $\theta_\tau = \pi - \theta_\ell$ (due to difference in definition) given in Ref.~\cite{Tanaka:2012nw} along with the above description, 
we can derive the SM contribution in Eqs.~\eqref{eq:fulldistribution} and \eqref{eq:angular}. 
Note that we have defined $H^{\pm}_{\pm} \equiv H_\pm (w)$ and $H^{0}_{0} \equiv H_0 (w)$ in the main text. 
The angular dependence for the case of the $V_2$ type operator is given simply by replacing $H_\pm (w) \to -C_{V_2} H_\mp (w)$ and $H_0 (w) \to -C_{V_2} H_0 (w)$.

As for the tensor NP operator, a similar procedure is applicable to obtain the angular distribution by taking 
\begin{align}
 \mathcal S^{\lambda_{\Dst}} (q^2,\theta_\ell) = {2G_F \over \sqrt{2}} V_{cb} C_T \sum_{\lambda,\lambda'} H^{\lambda_{\Dst}}_{\lambda,\lambda'} L_{\lambda,\lambda'}^{\lambda_\ell=+1/2} \,, 
\end{align}
where $L$ is again described in Ref.~\cite{Tanaka:2012nw}, $H^{\pm}_{\pm,0} = \pm H^{\pm}_{\pm,s} \equiv H^T_\pm (w)$, and $H^{0}_{+,-} = H^{0}_{0,s} \equiv H^T_0 (w)$. 
Since the lepton helicity of the tensor current is flipped compared with the SM current, one finds that the SM and tensor operators have no interference.

\section{Constraints from QCDSR}
\label{Ap:QCDSR}
The sub-leading IW functions, $\chi_{2,3}(w)$, $\eta(w)$, have been investigated by introducing QCDSR analysis up to two-loop perturbative corrections in the literature~\cite{Neubert:1992wq,Neubert:1992pn,Ligeti:1993hw}. 
In this approach, they are described as 
\begin{align}
 \chi_i(w) = [\alpha_s (1\text{GeV})]^{1/3}\, \bar\chi_i(w) \,, 
 \quad\quad
 \eta(w) = {1\over3} + \Delta(w) \,, 
 \label{Eq:QCDSRchieta}
\end{align}
with 
\begin{align}
 \bar\chi_2(w) \left[ F^2 \bar\Lambda e^{-2\bar\Lambda/T} \right] =\,- 
 & {\alpha_sT^4 \over 8\pi^3} \left( {2\over w+1} \right)^2 \left( {1-r(w) \over w-1} +2 \right) \delta_3 ({\omega_0 \over T}) \\
 & + {\alpha_s T  \langle \bar qq \rangle \over 6\pi} \left( {1-r(w) \over w-1} +{1 \over w+1} \right) \delta_0 ({\omega_0 \over T}) - {\langle \alpha_s GG \rangle \over 96\pi} \, {2\over w+1} \,, \notag \
  \label{Eq:QCDSRchi2}
\end{align}
\begin{align}
 \bar\chi_3(w) \left[ F^2 \bar\Lambda e^{-2\bar\Lambda/T} \right] =\, 
 & {\alpha_sT^4 \over 8\pi^3} \left( {2\over w+1} \right)^2 \left( w r(w) - 1 + \ln {w+1\over2} \right) \delta_3 ({\omega_0 \over T}) \\
 & + {3 \delta\omega_2 \over 32\pi^2} \omega_0^3 e^{-\omega_0/T} \left[ \left( {2\over w+1} \right)^2 -\xi(w) \right] \notag\\
 & + {\alpha_s T  \langle \bar qq \rangle \over 6\pi} \big[2 -r(w) -\xi(w) \big] \delta_0 ({\omega_0 \over T}) \notag\\
 & + {\langle \alpha_s GG \rangle \over 96\pi} \left[ {2\over w+1} -\xi(w) \right] - {\langle \bar q g_s \sigma_{\mu\nu} G^{\mu\nu} q \rangle \over 48T} \big[ 1 -\xi(w) \big] \,, \notag 
 \label{Eq:QCDSRchi3}
\end{align}
\begin{align}
 \Delta(w) \left[ \xi(w) F^2 \bar\Lambda e^{-2\bar\Lambda/T} \right] = \, 
 & {\alpha_sT^4 \over 12\pi^3} \left( {2\over w+1} \right)^2 \left( 11 + 6w + (3+w) r(w) \right) \delta_3 ({\omega_0 \over T}) \\
 & - {2\alpha_s T  \langle \bar qq \rangle \over 9\pi} \left(7 + (3-w)r(w) \right) \delta_0 ({\omega_0 \over T}) \notag\\
 & + {\langle \alpha_s GG \rangle \over 72\pi} \, {w-1\over w+1} + {\langle \bar q g_s \sigma_{\mu\nu} G^{\mu\nu} q \rangle \over 18T} (w-1) \,, \notag
  \label{Eq:QCDSReta}
\end{align}
where 
\begin{align}
 r(w) = {1 \over \sqrt{w^2-1}} \ln (w + \sqrt{w^2-1}) \,, 
 \quad\quad
 \delta_n (x) = {1 \over \Gamma (n+1)} \int_0^x dz z^n e^{-z} \,. 
\end{align}
The continuum threshold $\omega_0$ and Borel parameter $T$ control stability of the sum rule, as will be explained below. 
The renormalized factor $[\alpha_s (1\text{GeV})]^{1/3}$ connects the sub-leading IW functions in QCD $\bar\chi_i (w)$ to our basis $\chi_i (w)$.

The prefactors, presented with $\big[ \cdots \big]$ in Eqs.~\eqref{Eq:QCDSRchi2}--\eqref{Eq:QCDSReta}, contain the leading IW function $\xi(w)$, heavy meson decay constant $F$ (HQET basis), and heavy quark-meson mass difference $\bar\Lambda$. 
From two-current correlator, one finds 
\begin{align}
 \label{B}
 F^2 \bar\Lambda e^{-2\bar\Lambda/T} &= \frac{9 T^4}{8\pi^2}\delta_3\!\left(\frac{\omega_0}{T}\right)-\frac{\langle \bar q g_s \sigma_{\mu\nu} G^{\mu\nu} q \rangle}{4T}\,,  &\!\! (\text{see Ref.~\cite{Neubert:1992wq}}) \\[0.5em]
 \label{C}
 \xi(w) F^2 \bar\Lambda e^{-2\bar\Lambda/T} & =  \frac{9 T^4}{8\pi^2}\left(\frac{2}{1+w}\right)^2\delta_3\!\left(\frac{\omega_0}{T}\right)-\frac{2w+1}{3}\frac{\langle \bar q g_s \sigma_{\mu\nu} G^{\mu\nu} q \rangle}{4T}\,,  &\!\! (\text{see Ref.~\cite{Ligeti:1993hw}})
\end{align}
while $\xi(w)$ can be independently obtained as~\cite{Neubert:1992pn}
\begin{align}
 \xi(w) & = { K(T,\omega_0,w) \over K(T,\omega_0,1) }, \quad \left(\text{equivalently}\,~ \xi(w)F^2 e^{-2\bar\Lambda/T} = K(T,\omega_0,w)\,, \right)  \notag \\[0.5em]
 K(T,\omega_0,w)& = \frac{3T^3}{4\pi^2}\left(\frac{2}{1+w}\right)^2\delta_2\!\left(\frac{\omega_0}{T}\right) -\langle \bar{q} q \rangle+\frac{2w+1}{3}\frac{ \langle \bar q g_s \sigma_{\mu\nu} G^{\mu\nu} q \rangle }{4T^2} \,. 
 \label{Eq:QCDSRxi}
\end{align}
Note that $-{1\over2}{\partial \over \partial T^{-1}} K(T,\omega_0,w)$ equals to r.h.s. of Eq.~\eqref{C} and hence these expressions are consistent with each other.

Input parameters for the QCDSR predictions consist of the decay constant $F=(0.3\pm0.05) \text{GeV}^{3/2}$~\cite{Neubert:1992wq}, the mass difference $\bar\Lambda=(0.5\pm0.07)\text{GeV}$~\cite{Neubert:1992wq}, 
the spin-symmetry-violating correction $\delta\omega_2=(-0.1\pm0.02)\text{GeV}$ only for $\chi_3(w)$~\cite{Neubert:1992pn}, 
and the following vacuum condensates: 
\begin{align}
 \langle \bar qq \rangle & = \,-(0.25\pm0.01~ \rm{GeV} )^3\,, &(\text{from Refs.~\cite{Colangelo:2000dp,Ioffe:2005ym,Wang:2013ff}}) \\
 \langle \alpha_s GG \rangle & = \,(6.35\pm0.35)\times10^{-2}~\rm{GeV}^4\,, &(\text{from Ref.~\cite{Narison:2018dcr}}) \\
 \langle \bar q g_s \sigma_{\mu\nu} G^{\mu\nu} q \rangle & = \,m_0^2\langle \bar qq \rangle~~\text{with}~~m_0^2=(0.8\pm0.2)\text{GeV}^2 \,. &(\text{from Refs.~\cite{Belyaev:1982sa,Dosch:1997wb,DiGiacomo:2004ff}}) 
\end{align}
The continuum threshold $\omega_0$ and the Borel parameter $T$ have been determined in the literature so that $\bar\chi_{2,3}(1)$ and  $\Delta(1)$ are stabilized. 
In our case, concerning higher derivatives such as $\IWci{2}{2,3}$ and $\IWet{2}$, we take 
\begin{align}
 \label{Eq:controlparameter}
 0.7\,\text{GeV} < T < 1\,\text{GeV} \,, \quad 1.7\,\text{GeV} < \omega_0 < 2.3\,\text{GeV} \,. 
\end{align} 
Substituting QCDSR for $F$ and $\bar\Lambda$ as in Eqs.~\eqref{B} and \eqref{Eq:QCDSRxi}, and then taking numerical input within $1\sigma$ uncertainties, we obtain the constraints as in Eqs.~\eqref{Eq:QCDSR1}--\eqref{Eq:QCDSR3} of the main text. 
Note that $\hat\chi_i(w) = \chi(w) / \xi(w)$ and we take the conservative ranges for the uncertainties, which is in agreement with Ref.~\cite{Bernlochner:2017jka}.

\section{Fit results with some details}
\label{Ap:FitResult}
Here we write down useful output data obtained by our fit analyses. 
First, we show correlation among our fit results of the HQET parameters, for SM $(3/2/1)$ in Tables~\ref{Tab:fit:corrA1}--\ref{Tab:fit:corrB2}, and for SM $(2/1/0)$ in Tables~\ref{Tab:fit:corrAA1}--\ref{Tab:fit:corrBB2}.

We then provide our fit results of $\mathcal G(w)$ and $\mathcal F(w)$ for SM $(3/2/1)$ with the $z$ expansion forms as defined in Eqs.~\eqref{Eq:normalizationG} and \eqref{Eq:normalizationF}: 
\begin{align}
 \text{corr.}({\mathcal G})
 =
 \begin{pmatrix}
 1.0000 & -0.4626 & 0.2962 & -0.1886
 \\
 -0.4626 & 1.0000 & -0.7231 & 0.4812
 \\
 0.2962 & -0.7231 & 1.0000 & -0.9278
 \\
 -0.1886 & 0.4812 & -0.9278 & 1.0000
 \end{pmatrix}  \,, 
\end{align}
for $( \mathcal G(1), g_1, g_2, g_3 )$ and 
\begin{align}
 \text{corr.}({\mathcal F}^2)
 =
 \begin{pmatrix}
 1.0000 & -0.2661 & 0.1328 & -0.2465
 \\
 -0.2661 & 1.0000 & -0.8456 & 0.8270
 \\
 0.1328 & -0.8456 & 1.0000 & -0.8367
 \\
 -0.2465 & 0.8270 & -0.8367 & 1.0000
 \end{pmatrix}  \,, 
\end{align}
for $( \mathcal F(1), f_1^2, f_2^2, f_3^2 )$. 
In turn, $R_D$ -- $R_\Dst$ correlations are obtained as 
$-0.15$, $-0.03$, $-0.81$, and $-0.57$ for SM $(2/1/0)$, SM $(3/2/1)$, $\text{SM}+V_2$ $(2/1/0)$, and $\text{SM}+V_2$ $(3/2/1)$, respectively.

\begin{figure}[t!]
\begin{center}
\includegraphics[viewport=0 0 1024 768, width=34em]{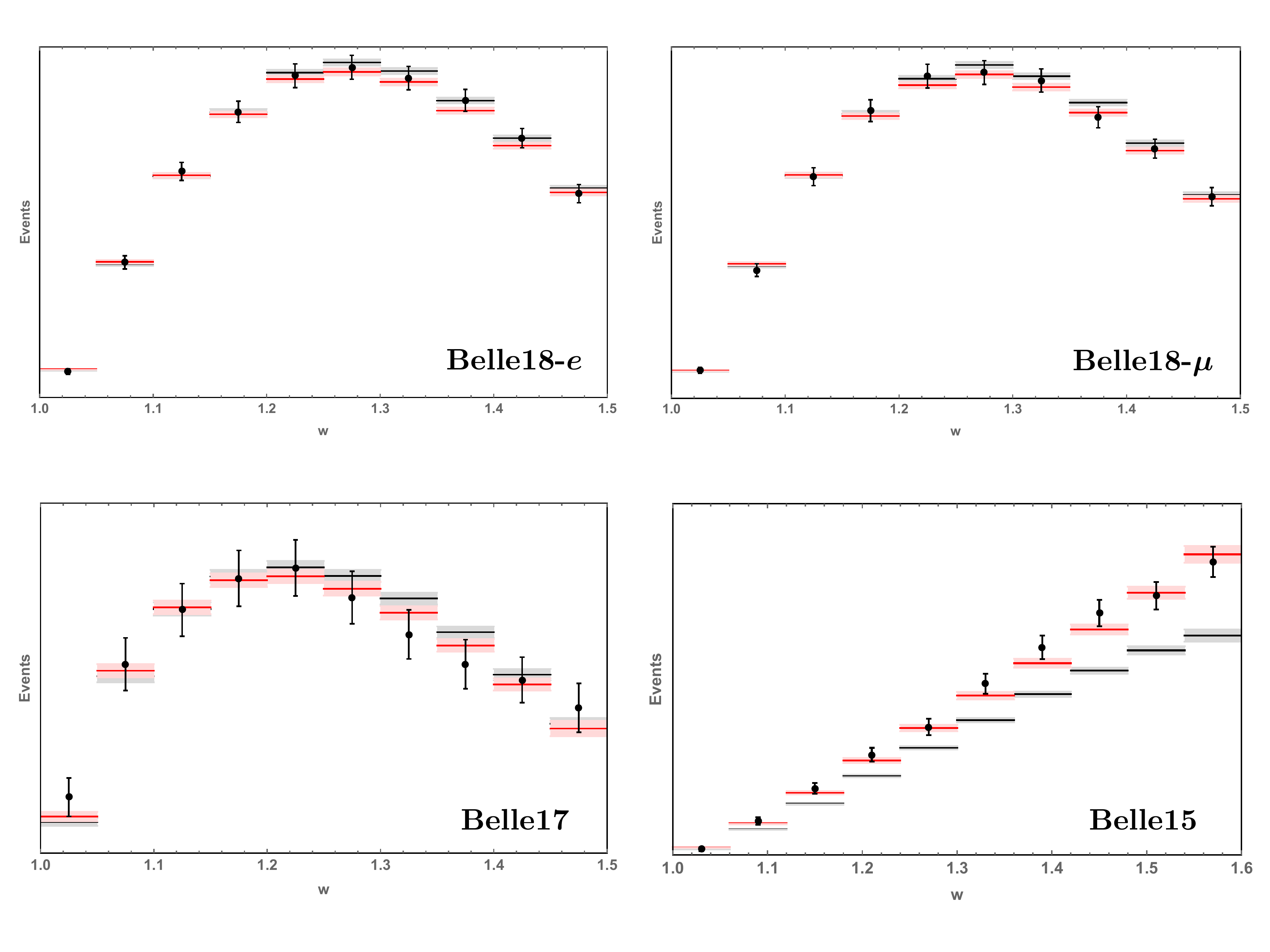} 
\caption{
 Binned decay distributions with respect to $w$ with the comparisons between data and the fit results from the SM $(2/1/0)$ [red] and SM $(3/2/\text{-})$ [gray] scenarios. 
} 
\label{Fig:distribution_w}
\end{center}
\end{figure}

\begin{figure}[t!]
\begin{center}
\includegraphics[viewport=0 0 1024 768, width=34em]{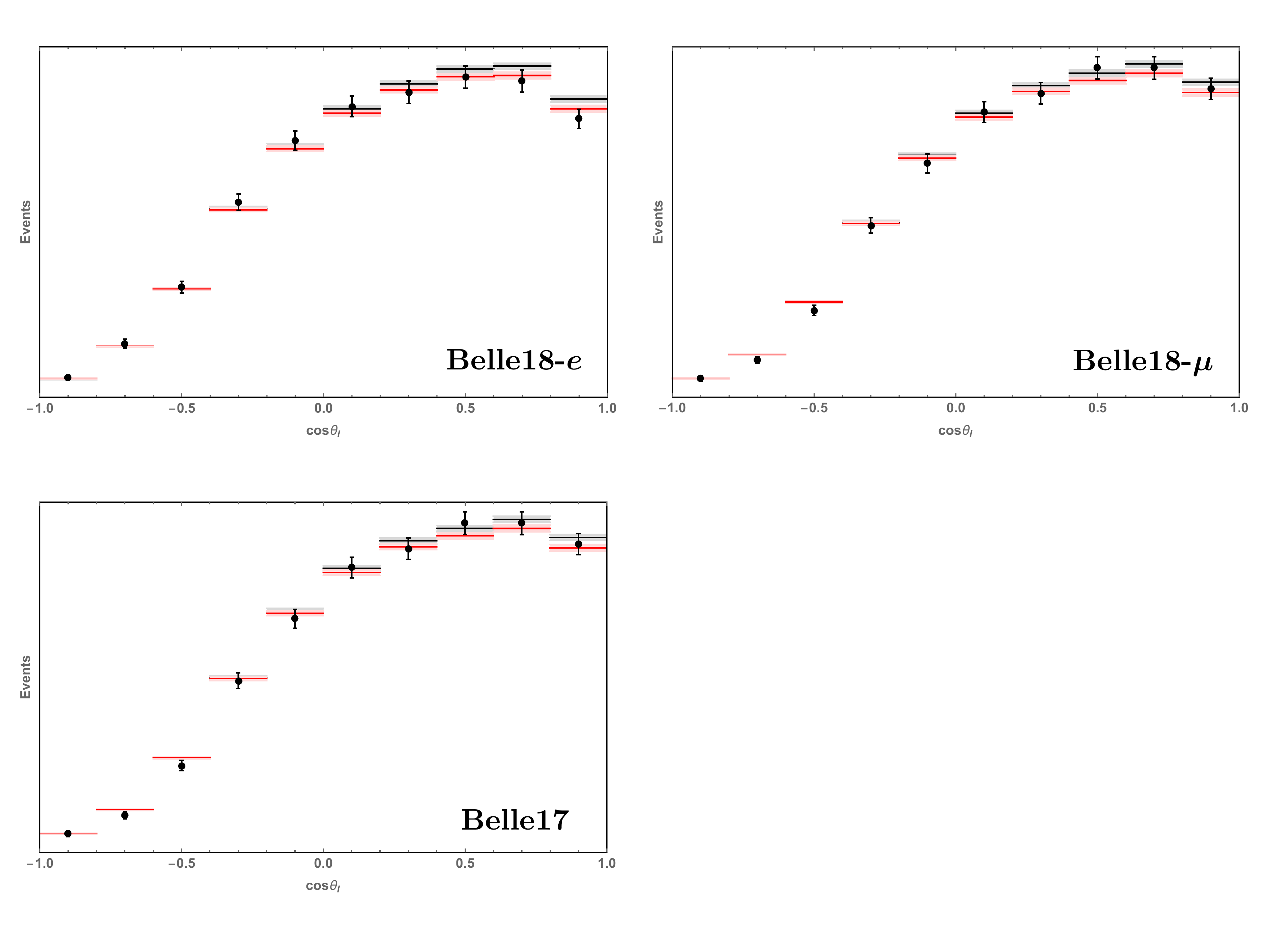} 
\caption{
 Binned decay distributions of $\cos\theta_\ell$. 
 Conventions are the same as in Fig.~\ref{Fig:distribution_w}. 
} 
\label{Fig:distribution_Thl}
\end{center}
\end{figure}

\begin{figure}[t!]
\begin{center}
\includegraphics[viewport=0 0 1024 768, width=34em]{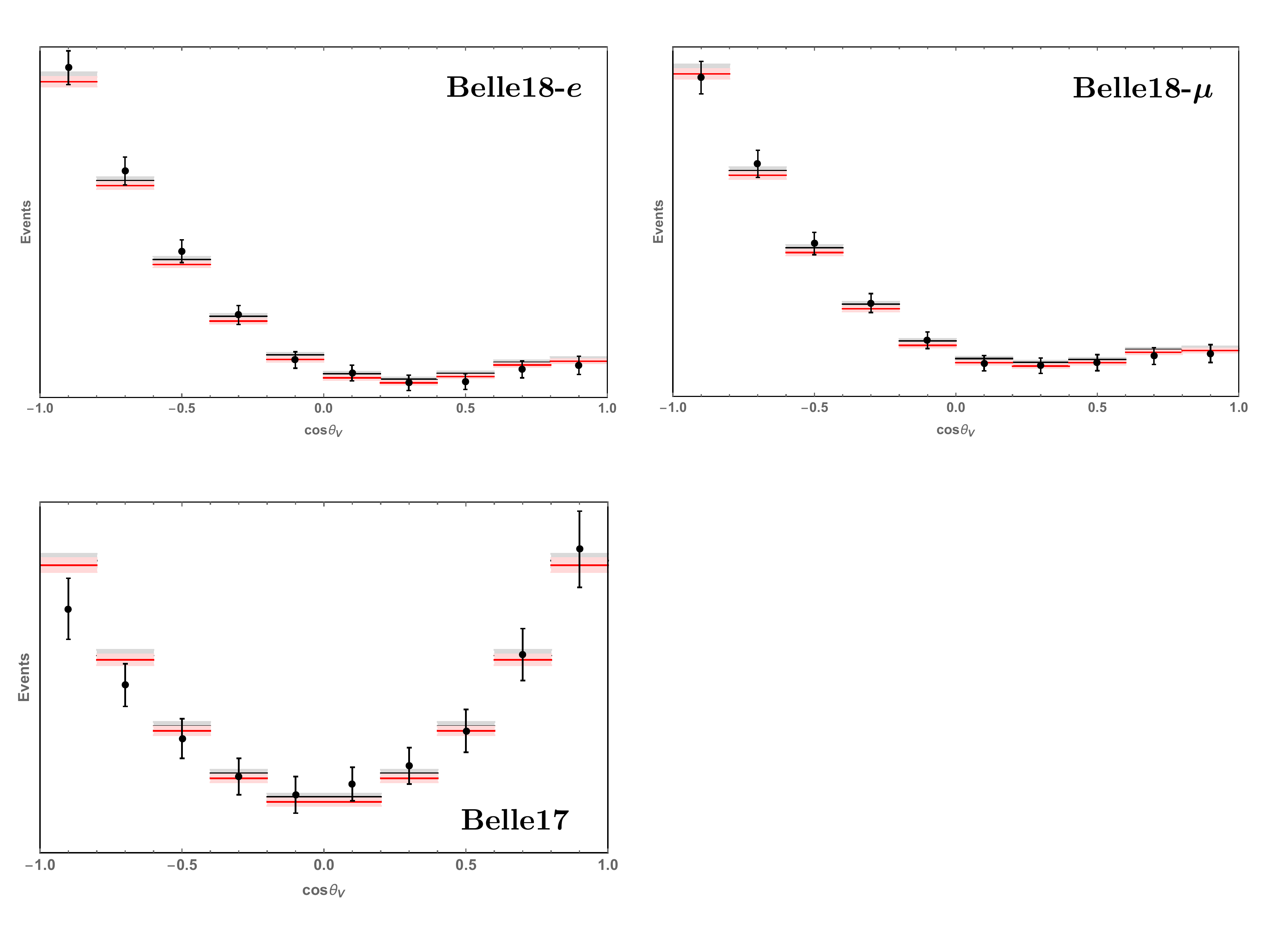} 
\caption{
 Binned decay distributions of $\cos\theta_V$. 
 Conventions are the same as in Fig.~\ref{Fig:distribution_w}. 
} 
\label{Fig:distribution_ThV}
\end{center}
\end{figure}

\begin{figure}[t!]
\begin{center}
\includegraphics[viewport=0 0 1024 768, width=34em]{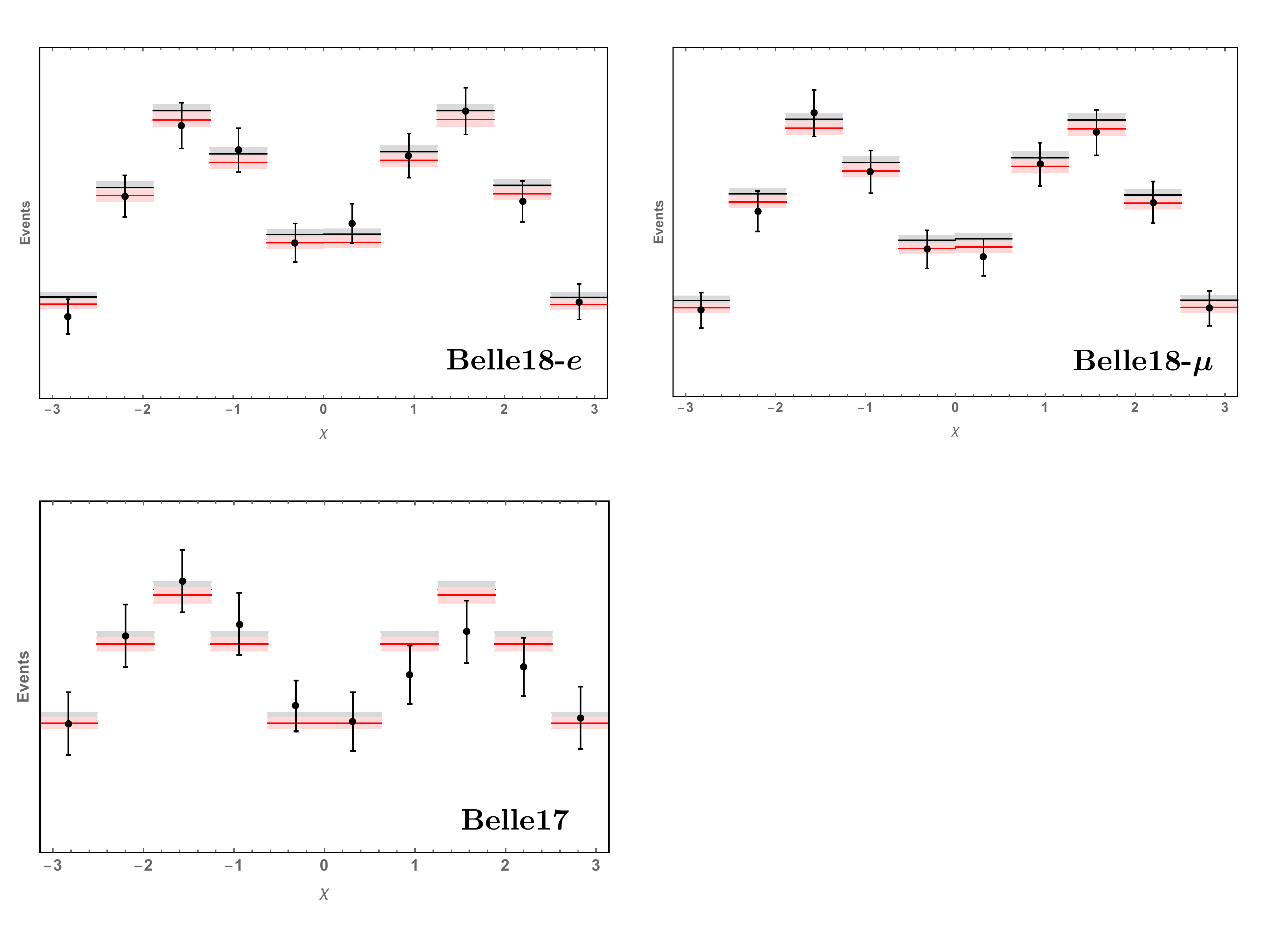} 
\caption{
 Binned decay distributions of $\chi$. 
 Conventions are the same as in Fig.~\ref{Fig:distribution_w}. 
} 
\label{Fig:distribution_Chi}
\end{center}
\end{figure}

In Figs.~\ref{Fig:distribution_w}--\ref{Fig:distribution_Chi}, we also show the binned decay distributions with respect to $(w,\cos\theta_\ell,\cos\theta_V,\chi)$ with the comparisons 
between data and the fit results in the SM $(2/1/0)$ [red] and SM $(3/2/\text{-})$ [gray] scenarios, in order to visualize the improvement on the fits. 
Note that the distributions for Belle17 (Belle18) are those of the decay rates (folded signal events) as explained in the main text.

\vspace{4em}

\begin{table}[h!]
\renewcommand{\arraystretch}{1.3}
  \begin{center}
  \scalebox{0.7}{
  \begin{tabular}{c|cccccccccccc}	
  \hline\hline
  corr. 		& $|V_{cb}|$ & $\IWxi{1}$  & $\IWxi{2}$  & $\IWxi{3}$  & $\IWci{0}{2}$  & $\IWci{1}{2}$  & $\IWci{2}{2}$  & $\IWci{1}{3}$  & $\IWci{2}{3}$  & $\IWet{0}$  & $\IWet{1}$  & $\IWet{2}$  \\
  \hline
  $|V_{cb}| $ 	&  $1.0000$ & $0.2807$ & $-0.2096$ & $0.1785$ & $0.0058$ & $-0.0621$ & $-0.0691$ & $0.0005$ & $0.3333$ & $0.0014$ & $-0.0078$ & $-0.0188$  \\
  $\IWxi{1}$  	&  $0.2807$ & $1.0000$ & $-0.9295$ & $0.8577$ & $0.1694$ & $-0.0862$ & $0.0192$ & $-0.0947$ & $0.3637$ & $0.2613$ & $0.0261$ & $0.0178$  \\
  $\IWxi{2}$ 	&  $-0.2096$ & $-0.9295$ & $1.0000$ & $-0.9851$ & $-0.1760$ & $0.0975$ & $-0.0187$ & $0.0263$ & $-0.3641$ & $-0.2324$ & $-0.0129$ & $-0.0160$ \\
  $\IWxi{3}$  	&  $0.1785$ & $0.8577$ & $-0.9851$ & $1.0000$ & $0.1718$ & $-0.0964$ & $0.0223$ & $0.0055$ & $0.3351$ & $0.2088$ & $0.0062$ & $0.0149$ \\
  $\IWci{0}{2}$  &  $0.0058$ & $0.1694$ & $-0.1760$ & $0.1718$ & $1.0000$ & $-0.0185$ & $-0.0183$ & $0.0484$ & $0.0573$ & $0.0518$ & $0.0123$ & $-0.0248$ \\
  $\IWci{1}{2}$  &  $-0.0621$ & $-0.0862$ & $0.0975$ & $-0.0964$ & $-0.0185$ & $1.0000$ & $-0.0089$ & $0.0772$ & $0.1369$ & $-0.0068$ & $-0.0258$ & $-0.0069$	   \\
  $\IWci{2}{2}$  &  $-0.0691$ & $0.0192$ & $-0.0187$ & $0.0223$ & $-0.0183$ & $-0.0089$ & $1.0000$ & $-0.0124$ & $0.3396$ & $0.0213$ & $-0.0065$ & $-0.0676$	 \\
  $\IWci{1}{3}$  &  $0.0005$ & $-0.0947$ & $0.0263$ & $0.0055$ & $0.0484$ & $0.0772$ & $-0.0124$ & $1.0000$ & $0.0404$ & $-0.0400$ & $0.0234$ & $0.0454$ \\
  $\IWci{2}{3}$  &  $0.3333$ & $0.3637$ & $-0.3641$ & $0.3351$ & $0.0573$ & $0.1369$ & $0.3396$ & $0.0404$ & $1.0000$ & $0.1322$ & $0.0241$ & $0.1332$  \\
  $\IWet{0}$  &  $0.0014$ & $0.2613$ & $-0.2324$ & $0.2088$ & $0.0518$ & $-0.0068$ & $0.0213$ & $-0.0400$ & $0.1322$ & $1.0000$ & $-0.0141$ & $-0.0440$  \\
  $\IWet{1}$  &  $-0.0078$ & $0.0261$ & $-0.0129$ & $0.0062$ & $0.0123$ & $-0.0258$ & $-0.0065$ & $0.0234$ & $0.0241$ & $-0.0141$ & $1.0000$ & $-0.0284$  \\
  $\IWet{2}$  & $-0.0188$ & $0.0178$ & $-0.0160$ & $0.0149$ & $-0.0248$ & $-0.0069$ & $-0.0676$ & $0.0454$ & $0.1332$ & $-0.0440$ & $-0.0284$ & $1.0000$  \\
  \hline\hline
  \end{tabular} 
  }
  \caption{Correlation among $\left\{ \IWxi{n},\, \IWci{n}{2,3},\, \IWet{n} \right\}$ -- $\left\{ \IWxi{n},\, \IWci{n}{2,3},\, \IWet{n} \right\}$ in SM $(3/2/1)$.    }
  \label{Tab:fit:corrA1}
  \end{center}
\end{table}

\begin{table}[h!]
\renewcommand{\arraystretch}{1.3}
  \begin{center}
  \scalebox{0.7}{
  \begin{tabular}{c|cccccccccccc}	
  \hline\hline
  corr. 		&  $\IWel{0}{1}$ & $\IWel{1}{1}$  & $\IWel{0}{2}$  & $\IWel{1}{2}$  & $\IWel{0}{3}$  & $\IWel{1}{3}$  & $\IWel{0}{4}$  & $\IWel{1}{4}$  & $\IWel{0}{5}$  & $\IWel{1}{5}$  & $\IWel{0}{6}$  & $\IWel{1}{6}$  \\
  \hline
  $|V_{cb}| $ 	&  $0.0143$ & $-0.4568$ & $-0.7995$ & $-0.3826$ & $-0.0109$ & $-0.0820$ & $0.0246$ & $0.0660$ & $0.3028$ & $-0.1854$ & $0.0598$ & $-0.1803$   \\
  $\IWxi{1}$  	&  $-0.3296$ & $-0.8549$ & $-0.3734$ & $-0.9066$ & $-0.0298$ & $-0.2690$ & $-0.2662$ & $0.0900$ & $0.0388$ & $-0.1974$ & $0.0921$ & $-0.5075$ \\
  $\IWxi{2}$ 	&  $0.3352$ & $0.7655$ & $0.3288$ & $0.7709$ & $0.0303$ & $0.2386$ & $0.2505$ & $-0.1322$ & $-0.0630$ & $0.2526$ & $-0.0827$ & $0.4896$ \\
  $\IWxi{3}$  	&  $-0.3226$ & $-0.6930$ & $-0.3006$ & $-0.6834$ & $-0.0304$ & $-0.2152$ & $-0.2314$ & $0.1472$ & $0.0664$ & $-0.2617$ & $0.0729$ & $-0.4596$  \\
  $\IWci{0}{2}$  &  $-0.0134$ & $0.0542$ & $-0.0332$ & $-0.0655$ & $-0.0019$ & $-0.0449$ & $-0.0507$ & $0.0232$ & $0.0009$ & $-0.0194$ & $0.0496$ & $-0.0113$  \\
  $\IWci{1}{2}$  &  $0.0032$ & $0.0622$ & $0.0642$ & $0.0880$ & $-0.0176$ & $-0.0209$ & $-0.0041$ & $-0.0177$ & $-0.0041$ & $0.0030$ & $-0.0217$ & $0.0338$    \\
  $\IWci{2}{2}$  &  $0.0102$ & $-0.0512$ & $0.0418$ & $0.0106$ & $0.0334$ & $-0.0758$ & $-0.0243$ & $-0.0036$ & $-0.0041$ & $0.0018$ & $0.0340$ & $-0.0372$ \\
  $\IWci{1}{3}$  &  $-0.3001$ & $-0.1324$ & $0.0115$ & $0.1256$ & $-0.0060$ & $0.0255$ & $0.0343$ & $0.0233$ & $0.0174$ & $-0.0224$ & $-0.0167$ & $0.0381$  \\
  $\IWci{2}{3}$  &  $0.0308$ & $-0.6003$ & $-0.3223$ & $-0.3126$ & $0.0186$ & $-0.1419$ & $-0.1426$ & $0.0792$ & $0.1145$ & $-0.1497$ & $0.0824$ & $-0.2139$   \\
  $\IWet{0}$  &  $-0.0999$ & $-0.1983$ & $-0.0952$ & $-0.1839$ & $0.0083$ & $0.0928$ & $-0.8842$ & $-0.5828$ & $-0.3765$ & $0.0521$ & $0.1033$ & $0.1541$ \\
  $\IWet{1}$  &  $0.0029$ & $-0.0396$ & $0.0008$ & $-0.0338$ & $-0.0111$ & $-0.0108$ & $0.0046$ & $-0.1132$ & $-0.0309$ & $0.0033$ & $-0.0085$ & $-0.0061$ \\
  $\IWet{2}$  &  $0.0418$ & $-0.1045$ & $0.0136$ & $-0.0221$ & $-0.0069$ & $-0.0516$ & $0.0687$ & $-0.2907$ & $0.0036$ & $-0.0178$ & $-0.0135$ & $-0.0342$  \\
  \hline\hline
  \end{tabular} 
  }
  \caption{Correlation among $\left\{ \IWxi{n},\, \IWci{n}{2,3},\, \IWet{n} \right\}$ -- $\left\{ \IWel{n}{1\text{-}6} \right\}$ in SM $(3/2/1)$.    }
  \end{center}
\end{table}

\begin{table}[h!]
\renewcommand{\arraystretch}{1.3}
  \begin{center}
  \scalebox{0.7}{
  \begin{tabular}{c|cccccccccccc}	
  \hline\hline
  corr. 		& $|V_{cb}|$ & $\IWxi{1}$  & $\IWxi{2}$  & $\IWxi{3}$  & $\IWci{0}{2}$  & $\IWci{1}{2}$  & $\IWci{2}{2}$  & $\IWci{1}{3}$  & $\IWci{2}{3}$  & $\IWet{0}$  & $\IWet{1}$  & $\IWet{2}$  \\
  \hline
  $\IWel{0}{1}$ 	&  $0.0143$ & $-0.3296$ & $0.3352$ & $-0.3226$ & $-0.0134$ & $0.0032$ & $0.0102$ & $-0.3001$ & $0.0308$ & $-0.0999$ & $0.0029$ & $0.0418$   \\
  $\IWel{1}{1}$  	&  $-0.4568$ & $-0.8549$ & $0.7655$ & $-0.6930$ & $0.0542$ & $0.0622$ & $-0.0512$ & $-0.1324$ & $-0.6003$ & $-0.1983$ & $-0.0396$ & $-0.1045$   \\
  $\IWel{0}{2}$ 	&  $-0.7995$ & $-0.3734$ & $0.3288$ & $-0.3006$ & $-0.0332$ & $0.0642$ & $0.0418$ & $0.0115$ & $-0.3223$ & $-0.0952$ & $0.0008$ & $0.0136$  \\
  $\IWel{1}{2}$  	&  $-0.3826$ & $-0.9066$ & $0.7709$ & $-0.6834$ & $-0.0655$ & $0.0880$ & $0.0106$ & $0.1256$ & $-0.3126$ & $-0.1839$ & $-0.0338$ & $-0.0221$  \\
  $\IWel{0}{3}$  &  $-0.0109$ & $-0.0298$ & $0.0303$ & $-0.0304$ & $-0.0019$ & $-0.0176$ & $0.0334$ & $-0.0060$ & $0.0186$ & $0.0083$ & $-0.0111$ & $-0.0069$ \\
  $\IWel{1}{3}$  &  $-0.0820$ & $-0.2690$ & $0.2386$ & $-0.2152$ & $-0.0449$ & $-0.0209$ & $-0.0758$ & $0.0255$ & $-0.1419$ & $0.0928$ & $-0.0108$ & $-0.0516$ 	   \\
  $\IWel{0}{4}$  &  $0.0246$ & $-0.2662$ & $0.2505$ & $-0.2314$ & $-0.0507$ & $-0.0041$ & $-0.0243$ & $0.0343$ & $-0.1426$ & $-0.8842$ & $0.0046$ & $0.0687$ 	 \\
  $\IWel{1}{4}$  &  $0.0660$ & $0.0900$ & $-0.1322$ & $0.1472$ & $0.0232$ & $-0.0177$ & $-0.0036$ & $0.0233$ & $0.0792$ & $-0.5828$ & $-0.1132$ & $-0.2907$  \\
  $\IWel{0}{5}$  &  $0.3028$ & $0.0388$ & $-0.0630$ & $0.0664$ & $0.0009$ & $-0.0041$ & $-0.0041$ & $0.0174$ & $0.1145$ & $-0.3765$ & $-0.0309$ & $0.0036$   \\
  $\IWel{1}{5}$  &  $-0.1854$ & $-0.1974$ & $0.2526$ & $-0.2617$ & $-0.0194$ & $0.0030$ & $0.0018$ & $-0.0224$ & $-0.1497$ & $0.0521$ & $0.0033$ & $-0.0178$  \\
  $\IWel{0}{6}$  &  $0.0598$ & $0.0921$ & $-0.0827$ & $0.0729$ & $0.0496$ & $-0.0217$ & $0.0340$ & $-0.0167$ & $0.0824$ & $0.1033$ & $-0.0085$ & $-0.0135$   \\
  $\IWel{1}{6}$  &  $-0.1803$ & $-0.5075$ & $0.4896$ & $-0.4596$ & $-0.0113$ & $0.0338$ & $-0.0372$ & $0.0381$ & $-0.2139$ & $0.1541$ & $-0.0061$ & $-0.0342$  \\
  \hline\hline
  \end{tabular} 
  }
  \caption{Correlation among $\left\{ \IWel{n}{1\text{-}6} \right\}$ -- $\left\{ \IWxi{n},\, \IWci{n}{2,3},\, \IWet{n} \right\}$ in SM $(3/2/1)$.    }
  \end{center}
\end{table}

\begin{table}[h!]
\renewcommand{\arraystretch}{1.3}
  \begin{center}
  \scalebox{0.7}{
  \begin{tabular}{c|cccccccccccc}	
  \hline\hline
  corr. 		&  $\IWel{0}{1}$ & $\IWel{1}{1}$  & $\IWel{0}{2}$  & $\IWel{1}{2}$  & $\IWel{0}{3}$  & $\IWel{1}{3}$  & $\IWel{0}{4}$  & $\IWel{1}{4}$  & $\IWel{0}{5}$  & $\IWel{1}{5}$  & $\IWel{0}{6}$  & $\IWel{1}{6}$  \\
  \hline
  $\IWel{0}{1}$ 	&  $1.0000$ & $0.2371$ & $0.0446$ & $0.2740$ & $0.0343$ & $0.0668$ & $0.3403$ & $-0.1619$ & $0.0205$ & $0.0542$ & $-0.0009$ & $0.1550$  \\
  $\IWel{1}{1}$  	&  $0.2371$ & $1.0000$ & $0.4762$ & $0.8234$ & $0.0194$ & $0.2486$ & $0.1900$ & $0.0097$ & $-0.1005$ & $0.1966$ & $-0.0863$ & $0.4662$  \\
  $\IWel{0}{2}$ 	&  $0.0446$ & $0.4762$ & $1.0000$ & $0.3309$ & $0.0141$ & $-0.0389$ & $0.0715$ & $-0.0389$ & $-0.0603$ & $0.0155$ & $-0.0258$ & $0.0183$   \\
  $\IWel{1}{2}$  	&  $0.2740$ & $0.8234$ & $0.3309$ & $1.0000$ & $0.0690$ & $0.3588$ & $0.1839$ & $-0.0939$ & $-0.0263$ & $0.3067$ & $-0.0342$ & $0.6221$   \\
  $\IWel{0}{3}$  &  $0.0343$ & $0.0194$ & $0.0141$ & $0.0690$ & $1.0000$ & $-0.5168$ & $-0.0007$ & $-0.0067$ & $0.0591$ & $0.0560$ & $0.9787$ & $-0.2958$  \\
  $\IWel{1}{3}$  &  $0.0668$ & $0.2486$ & $-0.0389$ & $0.3588$ & $-0.5168$ & $1.0000$ & $-0.0759$ & $-0.1192$ & $-0.1505$ & $0.3221$ & $-0.5437$ & $0.8919$  	   \\
  $\IWel{0}{4}$  &  $0.3403$ & $0.1900$ & $0.0715$ & $0.1839$ & $-0.0007$ & $-0.0759$ & $1.0000$ & $0.4221$ & $0.3323$ & $-0.0423$ & $-0.0880$ & $-0.1174$  	 \\
  $\IWel{1}{4}$  &  $-0.1619$ & $0.0097$ & $-0.0389$ & $-0.0939$ & $-0.0067$ & $-0.1192$ & $0.4221$ & $1.0000$ & $0.2675$ & $-0.1021$ & $-0.0360$ & $-0.2420$   \\
  $\IWel{0}{5}$  &  $0.0205$ & $-0.1005$ & $-0.0603$ & $-0.0263$ & $0.0591$ & $-0.1505$ & $0.3323$ & $0.2675$ & $1.0000$ & $-0.5833$ & $0.1354$ & $-0.2695$    \\
  $\IWel{1}{5}$  &  $0.0542$ & $0.1966$ & $0.0155$ & $0.3067$ & $0.0560$ & $0.3221$ & $-0.0423$ & $-0.1021$ & $-0.5833$ & $1.0000$ & $-0.0022$ & $0.5301$   \\
  $\IWel{0}{6}$  &  $-0.0009$ & $-0.0863$ & $-0.0258$ & $-0.0342$ & $0.9787$ & $-0.5437$ & $-0.0880$ & $-0.0360$ & $0.1354$ & $-0.0022$ & $1.0000$ & $-0.3591$    \\
  $\IWel{1}{6}$  &  $0.1550$ & $0.4662$ & $0.0183$ & $0.6221$ & $-0.2958$ & $0.8919$ & $-0.1174$ & $-0.2420$ & $-0.2695$ & $0.5301$ & $-0.3591$ & $1.0000$   \\
  \hline\hline
  \end{tabular} 
  }
  \caption{Correlation among $\left\{ \IWel{n}{1\text{-}6} \right\}$ -- $\left\{ \IWel{n}{1\text{-}6} \right\}$ in SM $(3/2/1)$.    }
  \label{Tab:fit:corrB2}
  \end{center}
\end{table}

\begin{table}[h!]
\renewcommand{\arraystretch}{1.3}
  \begin{center}
  \scalebox{0.7}{
  \begin{tabular}{c|ccccccccc}	
  \hline\hline
  corr. 		& $|V_{cb}|$ & $\IWxi{1}$  & $\IWxi{2}$  & $\IWci{0}{2}$  & $\IWci{1}{2}$  & $\IWci{1}{3}$  & $\IWet{0}$  & $\IWet{1}$  \\
  \hline
  $|V_{cb}| $ 	&  $1.0000$ & $0.1002$ & $0.1283$ & $-0.0494$ & $-0.0594$ & $-0.4822$ & $-0.2104$ & $-0.0876$  \\
  $\IWxi{1}$  	&  $0.1002$ & $1.0000$ & $-0.9012$ & $0.3099$ & $0.0193$ & $-0.1968$ & $-0.0321$ & $0.0926$   \\
  $\IWxi{2}$ 	&  $0.1283$ & $-0.9012$ & $1.0000$ & $-0.3127$ & $0.1381$ & $-0.0625$ & $0.0350$ & $-0.0639$  \\
  $\IWci{0}{2}$  &  $-0.0494$ & $0.3099$ & $-0.3127$ & $1.0000$ & $0.0962$ & $0.0604$ & $-0.2061$ & $0.0025$  \\
  $\IWci{1}{2}$  &  $-0.0594$ & $0.0193$ & $0.1381$ & $0.0962$ & $1.0000$ & $0.3050$ & $0.0147$ & $-0.1087$ 	   \\
  $\IWci{1}{3}$  &  $-0.4822$ & $-0.1968$ & $-0.0625$ & $0.0604$ & $0.3050$ & $1.0000$ & $-0.2991$ & $-0.0072$  \\
  $\IWet{0}$  &  $-0.2104$ & $-0.0321$ & $0.0350$ & $-0.2061$ & $0.0147$ & $-0.2991$ & $1.0000$ & $0.2309$   \\
  $\IWet{1}$  &  $-0.0876$ & $0.0926$ & $-0.0639$ & $0.0025$ & $-0.1087$ & $-0.0072$ & $0.2309$ & $1.0000$   \\
  \hline\hline
  \end{tabular} 
  }
  \caption{Correlation among $\left\{ \IWxi{n},\, \IWci{n}{2,3},\, \IWet{n} \right\}$ -- $\left\{ \IWxi{n},\, \IWci{n}{2,3},\, \IWet{n} \right\}$ in SM $(2/1/0)$.    }
  \label{Tab:fit:corrAA1}
  \end{center}
\end{table}

\begin{table}[h!]
\renewcommand{\arraystretch}{1.3}
  \begin{center}
  \scalebox{0.7}{
  \begin{tabular}{c|ccccccc}	
  \hline\hline
  corr. 		&  $\IWel{0}{1}$   & $\IWel{0}{2}$    & $\IWel{0}{3}$    & $\IWel{0}{4}$    & $\IWel{0}{5}$    & $\IWel{0}{6}$    \\
  \hline
  $|V_{cb}| $ 	&  $-0.0096$ & $-0.8276$ & $0.0225$ & $0.2246$ & $0.1837$ & $0.0306$   \\
  $\IWxi{1}$  	&  $-0.3082$ & $-0.4799$ & $0.0183$ & $0.0420$ & $0.3977$ & $0.1684$  \\
  $\IWxi{2}$ 	&  $0.3377$ & $0.2386$ & $-0.0229$ & $-0.0162$ & $-0.2336$ & $-0.1027$    \\
  $\IWci{0}{2}$  &  $0.0696$ & $-0.0604$ & $-0.0291$ & $0.1489$ & $0.0896$ & $0.0323$    \\
  $\IWci{1}{2}$  &  $-0.0916$ & $0.0459$ & $-0.1991$ & $0.0255$ & $0.1137$ & $-0.0848$	   \\
  $\IWci{1}{3}$  &  $-0.3307$ & $0.5291$ & $-0.1191$ & $0.2770$ & $-0.1426$ & $-0.1737$   \\
  $\IWet{0}$  &  $0.2129$ & $0.0913$ & $-0.0119$ & $-0.6829$ & $-0.3207$ & $0.0029$    \\
  $\IWet{1}$  &  $-0.1483$ & $0.0151$ & $-0.0307$ & $-0.3803$ & $-0.0484$ & $0.0398$    \\
  \hline\hline
  \end{tabular} 
  }
  \caption{Correlation among $\left\{ \IWxi{n},\, \IWci{n}{2,3},\, \IWet{n} \right\}$ -- $\left\{ \IWel{n}{1\text{-}6} \right\}$ in SM $(2/1/0)$.    }
  \label{Tab:fit:corrAA2}
  \end{center}
\end{table}

\begin{table}[h!]
\renewcommand{\arraystretch}{1.3}
  \begin{center}
  \scalebox{0.7}{
  \begin{tabular}{c|ccccccccc}	
  \hline\hline
  corr. 		& $|V_{cb}|$ & $\IWxi{1}$  & $\IWxi{2}$  & $\IWci{0}{2}$  & $\IWci{1}{2}$  & $\IWci{1}{3}$  & $\IWet{0}$  & $\IWet{1}$  \\
  \hline
  $\IWel{0}{1}$ & $-0.0096$ & $-0.3082$ & $0.3377$ & $0.0696$ & $-0.0916$ & $-0.3307$ & $0.2129$ & $-0.1483$ \\
  $\IWel{0}{2}$ & $-0.8276$ & $-0.4799$ & $0.2386$ & $-0.0604$ & $0.0459$ & $0.5291$ & $0.0913$ & $0.0151$ \\
  $\IWel{0}{3}$ & $0.0225$ & $0.0183$ & $-0.0229$ & $-0.0291$ & $-0.1991$ & $-0.1191$ & $-0.0119$ & $-0.0307$ \\
  $\IWel{0}{4}$ & $0.2246$ & $0.0420$ & $-0.0162$ & $0.1489$ & $0.0255$ & $0.2770$ & $-0.6829$ & $-0.3803$ \\
  $\IWel{0}{5}$ & $0.1837$ & $0.3977$ & $-0.2336$ & $0.0896$ & $0.1137$ & $-0.1426$ & $-0.3207$ & $-0.0484$ \\
  $\IWel{0}{6}$ & $0.0306$ & $0.1684$ & $-0.1027$ & $0.0323$ & $-0.0848$ & $-0.1737$ & $0.0029$ & $0.0398$ \\
  \hline\hline
  \end{tabular} 
  }
  \caption{Correlation among $\left\{ \IWxi{n},\, \IWci{n}{2,3},\, \IWet{n} \right\}$ -- $\left\{ \IWel{n}{1\text{-}6} \right\}$ in SM $(2/1/0)$.    }
  \label{Tab:fit:corrBB1}
  \end{center}
\end{table}

\begin{table}[h!]
\renewcommand{\arraystretch}{1.3}
  \begin{center}
  \scalebox{0.7}{
  \begin{tabular}{c|ccccccc}	
  \hline\hline
  corr. 		&  $\IWel{0}{1}$   & $\IWel{0}{2}$    & $\IWel{0}{3}$    & $\IWel{0}{4}$    & $\IWel{0}{5}$    & $\IWel{0}{6}$    \\
  \hline
  $\IWel{0}{1}$ &  $1.0000$ & $0.0968$ & $0.0519$ & $0.1446$ & $-0.1482$ & $0.0144$  \\
  $\IWel{0}{2}$ &  $0.0968$ & $1.0000$ & $0.0288$ & $-0.1439$ & $-0.0969$ & $0.0141$  \\
  $\IWel{0}{3}$ &  $0.0519$ & $0.0288$ & $1.0000$ & $0.0166$ & $0.1686$ & $0.9515$   \\
  $\IWel{0}{4}$ &  $0.1446$ & $-0.1439$ & $0.0166$ & $1.0000$ & $0.2297$ & $-0.0025$   \\
  $\IWel{0}{5}$ &  $-0.1482$ & $-0.0969$ & $0.1686$ & $0.2297$ & $1.0000$ & $0.4106$   \\
  $\IWel{0}{6}$ &  $0.0144$ & $0.0141$ & $0.9515$ & $-0.0025$ & $0.4106$ & $1.0000$   \\
  \hline\hline
  \end{tabular} 
  }
  \caption{Correlation among $\left\{ \IWel{n}{1\text{-}6} \right\}$ -- $\left\{ \IWel{n}{1\text{-}6} \right\}$ in SM $(2/1/0)$.    }
  \label{Tab:fit:corrBB2}
  \end{center}
\end{table}

\end{document}